\documentclass{article}

\usepackage{iclr2027_conference,times}

\usepackage{amsmath,amsfonts,bm}

\def\eqref#1{equation~\ref{#1}}

\def\1{\bm{1}}

\DeclareMathAlphabet{\mathsfit}{\encodingdefault}{\sfdefault}{m}{sl}
\SetMathAlphabet{\mathsfit}{bold}{\encodingdefault}{\sfdefault}{bx}{n}

\usepackage[utf8]{inputenc}
\usepackage[T1]{fontenc}
\usepackage{amsmath}
\usepackage{amsfonts}
\usepackage{nicefrac}
\usepackage{microtype}
\usepackage[table]{xcolor}
\usepackage{graphicx}
\graphicspath{{./}}
\usepackage{subcaption}
\usepackage{booktabs}
\usepackage{multirow}
\usepackage{tabularx}
\usepackage{array}
\usepackage{pifont}
\usepackage{adjustbox}
\usepackage{enumitem}
\usepackage{wrapfig}
\usepackage{hyperref}
\usepackage{url}
\usepackage{arydshln}

\ifdefined\pdfsuppresswarningpagegroup
\fi

\hypersetup{
  pdftitle={Unlocking Spatial Grounding in Large Audio-Visual Retrieval Models},
  pdfauthor={Hugo Malard, Michel Olvera, Sanjeel Parekh, Gael Richard, Slim Essid, Stephane Lathuiliere}
}

\newcommand{\cmark}{\ding{51}}%
\newcommand{\xmark}{\ding{55}}%
\newcommand{\tightpar}{\looseness=-1\relax}
\title{Unlocking Spatial Grounding in Large Audio-Visual Retrieval Models}

\author{
Hugo Malard$^{1}$
\And
Michel Olvera$^{1}$
\And
Sanjeel Parekh$^{2}$
\AND
Gaël Richard$^{1}$
\And
Slim Essid$^{3}$
\And
Stéphane Lathuilière$^{4}$
\AND
\normalfont
\makebox[\dimexpr\textwidth-2\tabcolsep\relax][c]{%
  \begin{tabular}{@{}c@{}}
  $^{1}$LTCI, Télécom Paris, Institut Polytechnique de Paris, France \\
  $^{2}$Meta, Reality Labs Research \\
  $^{3}$NVIDIA, France \\
  $^{4}$Inria at Université Grenoble Alpes, CNRS, LJK, France
  \end{tabular}%
}
}

\iclrfinalcopy

\begin{document}

\maketitle
\fancyhead{}
\renewcommand{\headrulewidth}{0pt}

\begin{abstract}
    \tightpar
    Weak supervision sets a practical regime for audio-visual sound source localization as dense spatial annotations are costly to obtain at scale. The task, however, remains challenging, as models must locate sound sources from temporally aligned audio-visual data without pixel-level supervision. Recent large-scale audio-visual retrieval models, trained at unprecedented scale, encode rich multimodal structure. We show their latent representations, though optimized for global alignment, can nonetheless enable fine-grained spatial grounding. While spatial detail is progressively lost in the upper layers of retrieval backbones due to global pooling, intermediate visual tokens retain highly structured spatial information. To exploit this, we introduce LAIP (\emph{Localization via Audio-Informed Pooling}), a framework that employs a lightweight \emph{Audio-informed Spatial Pooling} (AiSP) to replace the standard global aggregation module. By querying intermediate visual tokens with audio aligned at the frame level, LAIP recovers localized spatial information that is otherwise discarded by the retrieval pipeline, with the largest gains observed for PE-AV, a stack with underlying temporal aggregation. Our approach achieves state-of-the-art performance on AVSBench and AVATAR, nearly doubling previous results on the latter, improving average CIoU from 13.21 to 26.22.
\end{abstract}

\section{Introduction}

\tightpar Audio-visual sound source localization aims to identify, in each video frame, the spatial regions of active sound~\cite{EZ-VSL,CloserLook}. As a fundamental problem in audio-visual understanding, it underpins applications in video editing, robotics, and multimodal retrieval.
Yet, it is inherently challenging: in real-world scenes, visual content is often dense and cluttered, while sound sources are typically sparse. Consequently, models must pinpoint the regions responsible for the audio, such as \textit{a ringing phone on a cluttered desk filled with silent objects}, or \textit{a moving car honking in dense traffic}.
This sparsity makes the task technically difficult, as it requires precise spatial disambiguation under weak supervision (only having access to the audio associated with the video), and dense spatial annotations are rarely available. Obtaining such annotations for videos is burdensome: annotators must delineate sounding regions frame by frame across long clips.


\tightpar In parallel, large-scale contrastive pretraining has produced strong multimodal retrieval models across image--text~\citep{clip,siglip}, image--any-modality~\citep{imagebind}, and audio--visual settings~\citep{peav} by aligning global embeddings of those modalities over large dataset. These models are appealing for weakly supervised localization because they learn rich semantic structure from large collections of paired data, potentially avoiding the need to train a localizer from scratch on small, curated datasets.

\newcommand{\inlinefig}[1]{%
  \raisebox{-0.2\height}{\includegraphics[height=1.2em]{#1}}%
}

\begin{figure}[t]
  \centering
\includegraphics[width=0.72\linewidth]{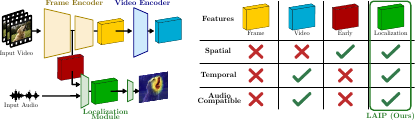}
  \caption{Global alignment suppresses spatial information in frame and video-level features. Earlier frame-encoder layers retain locality but are incompatible with the global multimodal space. LAIP learns locally grounded representations compatible with the audio space.}
  \label{fig:teaser}
\end{figure}

\tightpar However, this promise comes with challenges illustrated in Figure~\ref{fig:teaser}. Retrieval models are optimized for global alignment---matching audio to an entire frame or video---rather than for local matching of the specific region that emits the sound. It results in frame (\inlinefig{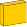}) and video features (~\inlinefig{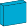}) with poor spatiality. In principle, dense spatial annotations could recover locality through fine-tuning, but collecting them is costly at scale. As a result, much of the prior work in the weakly supervised regime relies on localization-specific models trained directly for the task~\citep{fnac,EZ-VSL,AVATAR}. This leaves open a central question: \textit{can large pretrained retrieval models be adapted to recover fine-grained spatial grounding from such weak supervision alone?}

\tightpar Our hypothesis is that they can. Although global pooling progressively suppresses spatial detail in the upper layers of retrieval backbones, intermediate visual tokens still retain structured local information. This suggests intervening before the final aggregation stage (~\inlinefig{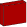})  to recover audio--visual correspondence from within a pretrained retrieval model, rather than learning localization entirely from scratch.

\tightpar To test this hypothesis, we instantiate it on PE-AV~\cite{peav}, a retrieval backbone composed of pretrained audio, frame, and video encoders. We propose Localization via Audio-Informed Pooling (LAIP), which inserts an \textit{Audio-informed Spatial Pooling (AiSP)} mechanism between the frame and video encoders: audio embeddings query dense visual tokens before temporal aggregation, yielding a sound-conditioned visual representation while remaining compatible with the original retrieval pipeline. 

\tightpar To train under this constraint, we adopt a SigLIP-style contrastive objective inspired by FLAIR~\cite{FLAIR}, which aligns the pooled video representation with its corresponding audio embedding while treating mismatched audio-conditioned views as negatives. This preserves cross-modal discrimination while remaining compatible with our pooled-token formulation. Although localization is not directly supervised, the attention maps induced by audio-informed spatial pooling provide strong spatial cues and lead to state-of-the-art sound source segmentation performance.

\tightpar At inference time, retrieval can still be performed using the standard PE-AV forward pass, as our approach reuses the same audio and visual representations for localization. However, this design changes the supervision interface: after AiSP, each frame is represented by a single pooled token passed to the video encoder, whereas standard localization methods such as EZ-VSL and its extensions require dense spatial tokens for region-level matching~\cite{EZ-VSL,AVATAR}.


Overall, our contributions are as follows:
\begin{itemize}[leftmargin=15pt, nosep, labelsep=5pt]
  \item \tightpar We show that large-scale audio--visual models pretrained for retrieval contain exploitable spatial information in intermediate visual tokens, making it possible to extend their capabilities to localization without dense spatial annotations.
  \item \tightpar We introduce LAIP, a framework that employs a lightweight audio-informed architecture to inject temporally aligned audio into the visual backbone prior to video pooling, enabling stronger cross-modal synergy while preserving compatibility with pretrained retrieval architectures. 
  \item \tightpar We show that, with the right architecture and regularization, global audio--video supervision is sufficient to learn rich spatial audio--visual representations without spatial annotations. Our audio-conditioned training strategy yields a new state of the art on the reference AVSBench and AVATAR benchmarks, almost doubling the performance of previous models on the latter.
\end{itemize}

\section{Related work}
\paragraph{Sound source localization}
\tightpar Audio-visual sound source localization aims to identify the visual region responsible for a sound. Weakly supervised methods show that synchronized video--audio pairs already provide a useful grounding signal, while also exposing recurring failures such as attending to correlated but silent objects or mishandling off-screen sources \cite{Senocak2018,EZ-VSL,CloserLook,SSLTIE,marginnce,fnac,CrossModalAlignment}. The task has since expanded to segmentation-oriented and open-vocabulary settings \cite{AVSBench,AVSAM,lee2025audio,guo2024open,taco}.

\tightpar EZ-VSL \cite{EZ-VSL} remains a key reference point, formulating localization as multiple-instance learning. Later works revisited protocols \cite{CloserLook}, reduced false negatives \cite{fnac}, or strengthened semantic alignment \cite{CrossModalAlignment}. However, most methods remain frame-centric, aligning one audio representation with one image. This is restrictive for realistic videos, where sources move, appear off-screen, or switch across objects. AVATAR \cite{AVATAR} addresses this gap with a video-centric benchmark and TAVLO, which combines spatial and temporal attention under a temporal extension of the EZ-VSL objective.

\tightpar Our work is complementary to these localization-specific architectures. Instead of preserving spatial outputs throughout the model, we start from a retrieval encoder that compresses each frame or clip into pooled tokens before alignment, and ask whether localization can be recovered from this interface. PE-AV~\cite{peav} provides a concrete instance of this setting.

\paragraph{Large-scale learning of multimodal representations}
\tightpar Recent multimodal representation learning follows a scaling paradigm in which large encoders are trained on massive paired datasets to align modalities in a shared embedding space. CLIP \cite{clip} established this recipe for image--text learning, SigLIP~\cite{siglip} replaced the softmax contrastive loss with a pairwise sigmoid loss, ImageBind~\cite{imagebind} extended the shared-embedding perspective to additional modalities, and PE-AV~\cite{peav} brought the same philosophy to joint audio, video, and text learning using roughly 100 million curated audio--video pairs. However, these models are optimized for global retrieval: they align modalities through pooled representations rather than dense spatial correspondences, and therefore do not directly expose the spatial outputs required for sound source localization.

\tightpar FLAIR~\cite{FLAIR} addresses an analogous problem in image--text learning by using the text embedding to pool local visual tokens and recover text-conditioned image features from a globally trained model. Our setting is the audio-visual analogue: can audio pool local visual tokens so that a retrieval model becomes spatially informative?


\section{Method}
\subsection{Problem formulation and overview}
\label{sec:pbForm}
\tightpar Audio-visual sound source localization (AV-SSL) involves identifying the spatial regions within a video frame $x_t$ that correspond to the synchronized audio signal at time $t$. 
We formulate this task as the extraction of a per-frame spatial probability map over the visual tokens of $x_t$ to isolate the sounding sources. 
We build on top of a large-scale audio-visual retrieval model composed of three transformer-based encoders: an audio encoder, an image frame encoder, and a video encoder. PE-AV is particularly relevant for our setup as it contains very rich pretrained audio-visual semantics, but it exhibits a clear architectural bottleneck: dense frame-level visual tokens are ultimately compressed into a single token before temporal modeling. Our goal is to show that this retrieval architecture can be adapted into an audio source localizer through a principled modification of that interface.
Let $c_t^{L} \in \mathbb{R}^{F}$ denote the classification token [CLS] from the final layer $L$ of the frame encoder, $F$ being its dimension. 
The PE-AV architecture constructs a video-level representation by processing the sequence of frame-level [CLS] tokens $\{c_t^{L}\}_{t=1}^{T}$ through the video encoder. In parallel, the audio encoder produces a sequence of audio tokens $\{\tilde{a}_r\}_{r=1}^{R}$ at its native temporal resolution.
Consequently, downstream temporal modeling operates exclusively on a single aggregated token per frame, rather than the underlying spatial grid. 
While this \textit{[CLS]-based} design is effective for global retrieval tasks, it introduces an inherent information bottleneck for AV-SSL, as the video encoder is disconnected from the localized visual features necessary for precise spatial grounding.

\begin{figure}[t]
    \centering
    \includegraphics[width=0.85\linewidth]{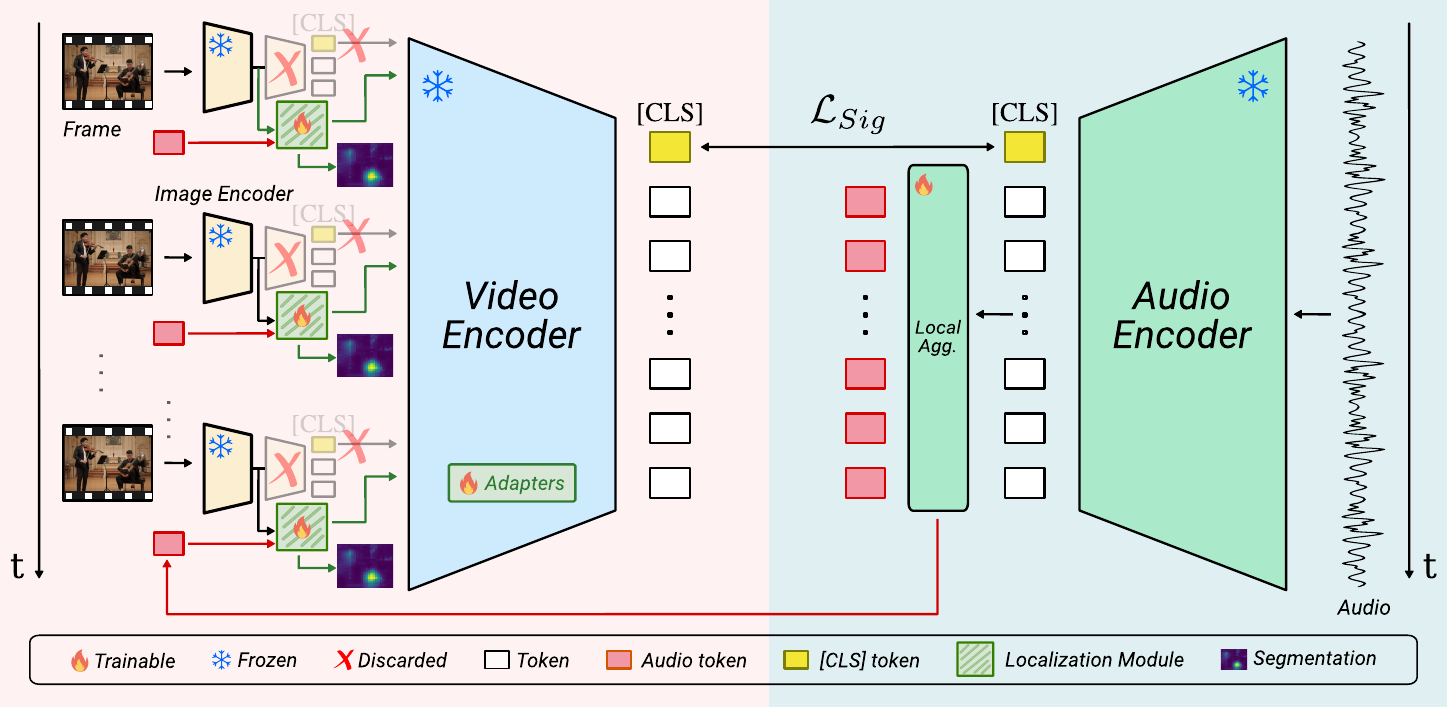}
    \caption{\textbf{LAIP: Localization via Audio-Informed Pooling}. The \textcolor[HTML]{ef5350}{\textbf{audio tokens}} are first forwarded to a \textcolor[HTML]{78C89F}{\textbf{local aggregator}} to match visual frame rate, and then used to pool the frame representation through our proposed \textcolor[HTML]{9FB39C}{\textbf{localization module}}. The pooled representations are used as input to the \textcolor[HTML]{1E90FF}{\textbf{video encoder}}, which outputs the \textcolor[HTML]{edde0e}{\textbf{final representation}} used to compute the sigmoid loss.}
    \label{fig:pipeline}
\end{figure}

\tightpar Consistent with the observations in \cite{PE}, we posit that spatial detail is progressively lost across the upper layers of the frame encoder, so that the final representation no longer retains the local information required for precise grounding. We therefore perform sound source localization from an intermediate layer $\ell$ of the frame encoder. Let $V_t^{\ell} \in \mathbb{R}^{HW \times F}$ denote the corresponding spatial tokens for frame $t$; in practice, we use the $16$th layer, i.e., eight layers before the final 24th layer.
Crucially, the audio and image-frame encoders operate at different temporal resolutions. We therefore transform the native audio tokens $\{\tilde{a}_r\}_{r=1}^{R}$ into frame-aligned representations using a lightweight local aggregator (a two-layer 1D convolutional network followed by temporal average pooling), which yields a synchronized audio embedding $a_t \in \mathbb{R}^{F}$ for each visual frame $x_t$.
The proposed method, is illustrated in Figure~\ref{fig:pipeline}. 

\subsection{Localization module}

\begin{figure}[ht]
  \centering
  \begin{subfigure}[t]{0.4\textwidth}
    \centering
    \includegraphics[height=3.6cm]{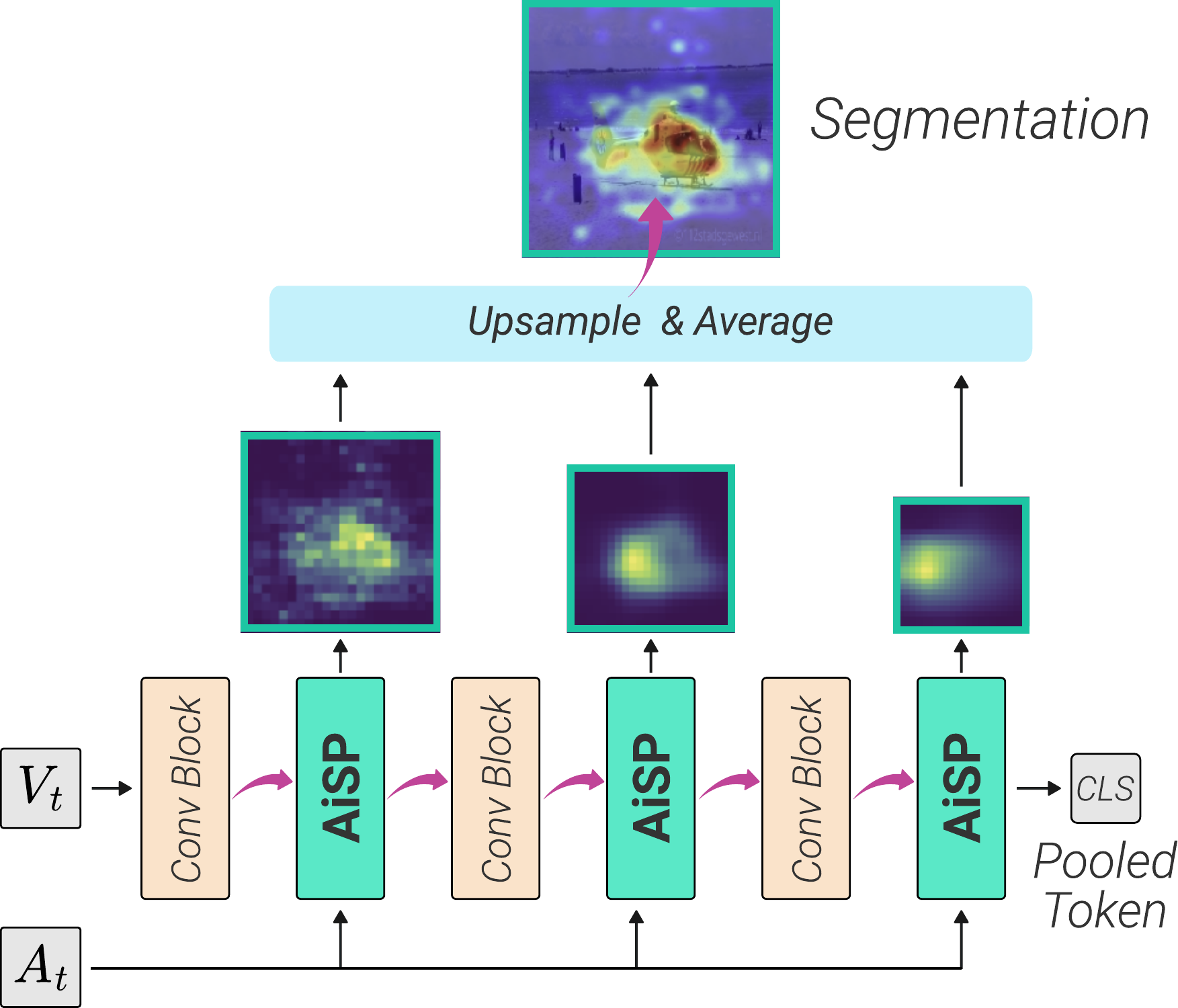}
    \caption{Localization module.}
    \label{fig:localizeModule}
  \end{subfigure}
  \hfill
  \begin{subfigure}[t]{0.58\textwidth}
    \centering
    \includegraphics[height=3.4cm]{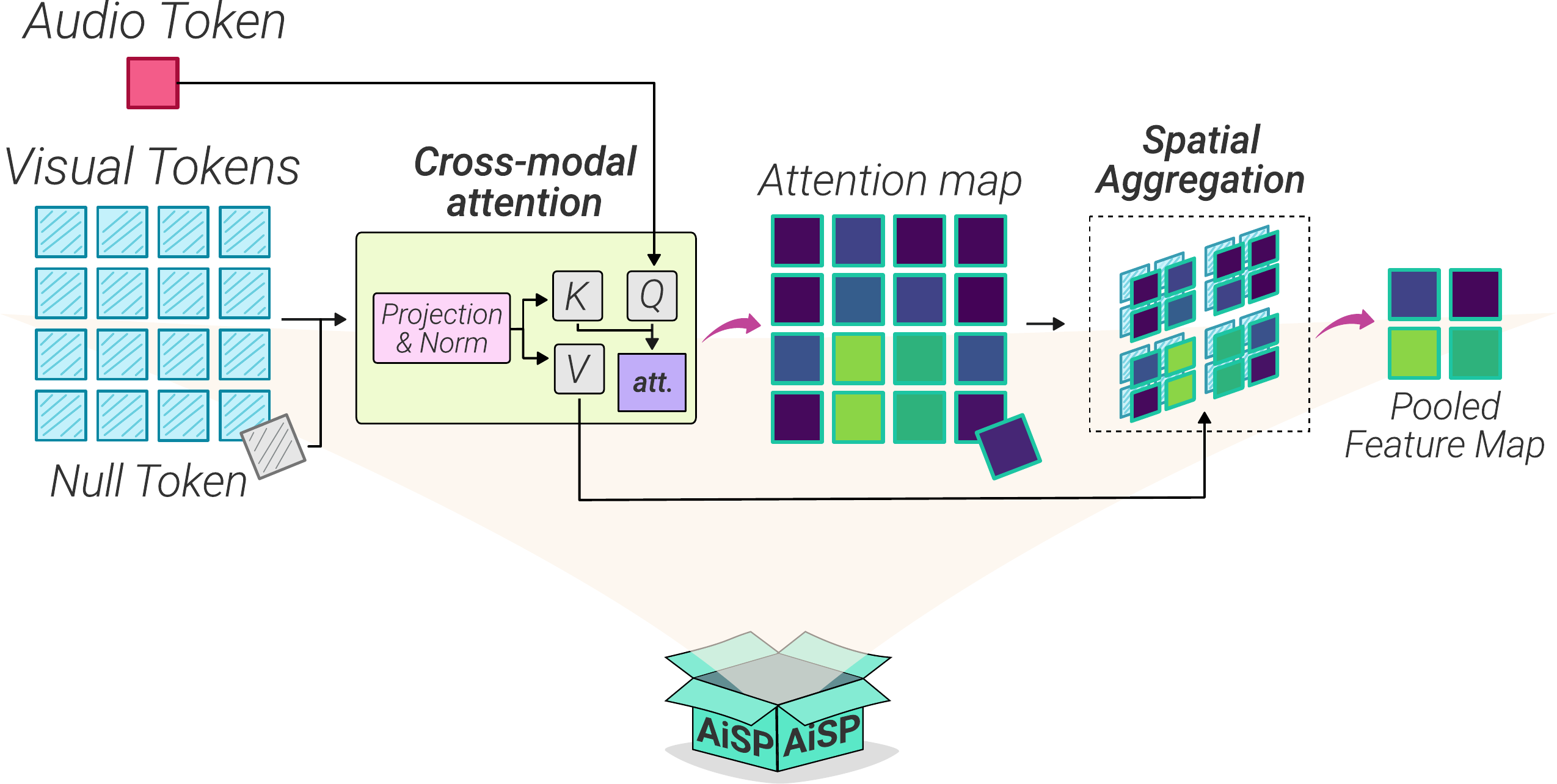}
    \caption{AiSP module.}
    \label{fig:aisp}
  \end{subfigure}

  \caption{Overview of our localization and AiSP modules, with $K=2$. Starting from PE visual tokens extracted from the image branch at the $16$-th frame-encoder layer, a frame-aligned audio token progressively pools the spatial grid into one sound-conditioned visual token per frame.}
  \label{fig:audioPooling}
\end{figure}

\tightpar Rather than attaching a separate localization head on top of frozen PE-AV features, LAIP modifies the interface between the frame encoder and the video encoder, by adding a localization module (illustrated in Figure~\ref{fig:localizeModule}). Its core building block is the Audio-informed Spatial Pooling (AiSP) modules, which progressively pool the intermediate visual tokens into an audio-informed [CLS] token $\tilde{c}_t^L$ while exposing attention maps for localization. This compatibility with the original PE-AV video encoder acts as a form of \textit{implicit distillation}: the model is encouraged to extract localized spatial information while still producing per-frame embeddings that the pretrained temporal stack can consume.

\tightpar A straightforward approach would be to apply a single cross-attention layer between the audio representation and the full set of spatial visual tokens. However, large-scale vision transformers are known to produce \textit{register tokens}, i.e., outlier tokens with disproportionately high norms and overly globalized semantics \cite{registers}, which can dominate cross-attention weights and degrade spatial precision. To circumvent this, LAIP adopts a hierarchical design inspired by the classic convolution-pooling paradigm \cite{lecun98}: we stack three AiSP modules, each followed by a lightweight convolutional adaptation block, and progressively reduce the spatial resolution of the visual tokens until they collapse into the replacement token $\tilde{c}_t^L$.

\tightpar This design keeps the pooling process spatially grounded, as the audio signal selectively attends to tokens corresponding to active sounding regions rather than to registers or tokens that have already aggregated global context. It also yields multi-scale attention maps across the successive pooling stages. At inference time, we combine these maps to obtain the final localization.


\paragraph{Audio-informed Spatial Pooling (AiSP)}
\tightpar This pooling layer (depicted in Figure~\ref{fig:aisp}) is designed with two objectives: (i) reducing the spatial resolution of the visual feature maps, and (ii) suppressing regions corresponding to non-sounding objects. To this end, we first estimate an attention map that localizes the sounding object at the current resolution, and then perform pooling conditioned on this localization to refine the features passed to subsequent layers. Since the model is trained through a cross-modal global objective, the underlying intuition is that improved localization of sounding objects yields more informative visual features for the temporal transformer. Consequently, the architecture is implicitly encouraged to produce accurate localization, leading to better alignment between visual and audio representations.

\tightpar Concretely, the AiSP operator down-samples the spatial token map by a factor of $K$ along each spatial dimension. Using the notation from Section~\ref{sec:pbForm}, we consider the frame-level visual tokens as an $H \times W$ grid, denoted by \(V_t \in \mathbb{R}^{H \times W \times F}\) for readability, and query them with the temporally aligned audio token \(a_t\).

\tightpar We then flatten \(V_t\) into a sequence of \(HW\) spatial tokens. Ignoring for now the optional null token introduced below, keys and values are obtained by linear projection and normalization, yielding \(\{k_i, v_i\}_{i=1}^{HW}\), while the query is obtained from the linearly projected and normalized audio token $a_t$ as \(q_a \in \mathbb{R}^F\). We then compute an attention matrix as
\[
\alpha_i = \frac{\exp\left(\frac{q_a^\top k_i}{\sqrt{F}}\right)}{\sum_{j=1}^{HW} \exp\left(\frac{q_a^\top k_j}{\sqrt{F}}\right)}, \quad i \in \{1,\dots,HW\},
\]
such that \(\sum_{i=1}^{HW} \alpha_i = 1\). Furthermore, as $\alpha$ reflects the image regions attended to by the audio modality, we employ it as a segmentation map.

\tightpar The spatial attention weights \(\{\alpha_{h,w}\}\) are then reshaped into an attention map \(M \in \mathbb{R}^{H \times W}\), preserving the two-dimensional structure of the frame. For each non-overlapping \(K \times K\) spatial block \(\mathcal{B}_{u,v}\), we define the pooled feature as:
\[
Y_{u,v} = \sum_{(h,w)\in\mathcal{B}_{u,v}} \alpha_{h,w}\,v_{h,w},
\]
where \(v_{h,w}\) denotes the projected and normalized visual token at location \((h,w)\). This yields a pooled feature map \(Y \in \mathbb{R}^{\frac{H}{K}\times \frac{W}{K}\times F}\). Unlike locally normalized attention pooling, this formulation preserves globally normalized cross-attention scores while restricting the aggregation to local spatial neighborhoods (but does not preserve the mass in each spatial block, as some of them can contain no audio-related information at all). Overall, this defines $P_K(V_t,a_t): \mathbb{R}^{H \times W \times F} \rightarrow \mathbb{R}^{\frac{H}{K} \times \frac{W}{K} \times F}$.

\tightpar Inspired by FLAIR, we propose appending a learnable token, \(x_{\emptyset}\), to the key tokens, before the pooling. It gives the audio query an explicit fallback option, allowing it to attend away from the image tokens when none of them is relevant. This is particularly useful in the audio-visual setting, where the audio content may differ substantially from the visible scene.




\subsection{Training and inference with global supervision}
\label{sec:training}

\tightpar Let $O^{(i,j)} \in \mathbb{R}^{F}$ denote the output of the video encoder when the visual tokens of video $i$ are aggregated by LAIP using the temporal tokens of audio $j$ as queries. Following \cite{FLAIR}, we train all $B^2$ audio--video combinations with a sigmoid-based contrastive objective:
\begin{equation}
  \mathcal{L}_{\mathrm{SIG}}
  = -\frac{1}{B^{2}}\sum_{i=1}^{B}\sum_{j=1}^{B}
  \log \sigma\!\left(
  y_{ij}\left[\tau\,\left\langle O^{(i,j)},A_g^{(j)}\right\rangle+b\right]
  \right).
\end{equation}
\tightpar Here, $y_{ij}=+1$ when $i=j$ and $y_{ij}=-1$ otherwise. Moreover, $\sigma$ denotes the sigmoid function, $\tau$ is a learnable logit scale, $b$ is a learnable bias, and $\langle\cdot,\cdot\rangle$ is the inner product. The visual pooling that produces $O^{(i,j)}$ is conditioned by the frame-aligned tokens $\{a_t^{(j)}\}_{t=1}^{T}$, while $O^{(i,j)}$ is compared with the global representation $A_g^{(j)}$ produced by the same audio encoder. The pair is positive only when audio $j$ is associated with video $i$.

\tightpar This objective preserves global discriminative power while specifically regularizing the sound-conditioned pooling mechanism that generates $O^{(i,j)}$. As in FLAIR, a query from one modality first conditions the visual stream before global contrastive matching; here the conditioning signal is audio rather than text, which distinguishes our loss from the standard SigLIP objective \cite{siglip} used in PE-AV.

\tightpar While $\mathcal{L}_{\mathrm{SIG}}$ supervises the final sound-conditioned video representation, it does not directly constrain the intermediate attention maps produced by the three-stage AiSP hierarchy. We therefore enforce consistency across resolutions. For frame $t$, the first two single-head stages produce maps $M_{1,t}$ and $M_{2,t}$, while the final, strongest-reduction stage produces $8$ maps $M_{3,t}^h$. The per-frame loss is:
\begin{equation}
  \mathcal{L}_{\mathrm{MRes}}^{t}
  = \left\lVert M_{1,t} - \operatorname{Up}(M_{2,t}) \right\rVert_{F}
  + \min_h \left\lVert M_{2,t} - \operatorname{Up}(M_{3,t}^h) \right\rVert_F,
\end{equation}
where $\operatorname{Up}$ is up-sampling and $\lVert\cdot\rVert_F$ is the Frobenius norm. The training loss $\mathcal{L}_{\mathrm{MRes}}$ averages $\mathcal{L}_{\mathrm{MRes}}^{t}$ over the frames and batch. At inference, the same multi-resolution criterion selects the final-stage head for each frame:
\begin{equation}
  h_t^{\star}
  = \operatorname*{arg\,min}_h
  \left\lVert M_{2,t} - \operatorname{Up}(M_{3,t}^h) \right\rVert_F,
\end{equation}
then produce the final heatmap for frame $t$ by averaging $M_{1,t}$, $\operatorname{Up}(M_{2,t})$, and $\operatorname{Up}(M_{3,t}^{h_t^{\star}})$ at a common resolution.

\tightpar During preliminary experiments, we observed that the model tended to allocate excessive attention to the learnable null token \(x_{\emptyset}\), particularly for negative audio-video pairs (more details in Appendix \ref{sec:hardNeg}). Because the null token participates in the attention normalization but is excluded from the pooled output, high attention to the null token reduces the total weight assigned to visual tokens. Accumulated across successive AiSP stages, this effect can drive the conditioned representation \(O^{(i,j)}\) toward zero and degrade its quality. We therefore directly penalize null attention. Let \(\alpha_{\emptyset,s}^{i,j,t}\) denote the attention assigned to \(x_{\emptyset}\) at stage \(s\) and frame \(t\) for video \(i\) queried with audio \(j\). We penalize its accumulation across all negative pairs, stages, and frames as follows.
$\mathcal{L}_{\emptyset}
  = \frac{1}{ST}
  \sum_{\substack{i \neq j \\
                     s, t}}
  \alpha_{\emptyset,s}^{i,j,t}$,
and incorporate it in the final objective:
\begin{equation}
  \mathcal{L}
  = \mathcal{L}_{\mathrm{SIG}}
  + \lambda \mathcal{L}_{\mathrm{MRes}}
  + \mu \mathcal{L}_{\emptyset}.
\end{equation}
This anti-collapse regularizer forces the negative branch to retain visual evidence, with $\mu$ and $\lambda$ controlling the strength of the null-attention and multi-resolution regularizers, respectively.
\tightpar Finally, small adapters \citep{adapters} bridge the difference between the pretrained frame token $c_t^L$ and our audio-informed $\tilde{c}_t^L$; all other pretrained weights remain frozen. LAIP preserves PE-AV's original retrieval forward pass: for retrieval, it bypasses the local audio aggregator, AiSP modules, and adapters, and feeds the original final-layer tokens $\{c_t^L\}_{t=1}^{T}$ to the unchanged frozen video encoder. Because this path has exactly the same operations and weights as PE-AV, its retrieval embeddings and scores are identical. The audio-conditioned path is used only for localization.
\section{Experiments}
\subsection{Datasets and implementation details}
\tightpar We evaluate LAIP on three benchmarks. Our primary benchmark is AVATAR~\citep{AVATAR}, a video-centric audio-visual localization dataset designed to capture temporal dynamics beyond frame-centric AVL settings. AVATAR contains 5{,}000 videos with an average duration of 10 seconds and 24{,}266 annotated frames across 80 target categories, with instance-segmentation annotations covering four scenarios: single-sound, mixed-sound, multi-entity, and off-screen. 

\tightpar We also evaluate on AVSBench~\citep{AVSBench}, a benchmark for sound-prompted audio-visual segmentation and localization that provides pixel-level masks for sounding objects. We report results on its single-source (S4) and multi-source (MS3) test sets.
Finally, we also evaluate our method on ADE-SP, the sound-prompted ADE20k benchmark introduced in~\citep{denseAV}, which pairs ADE20k images and segmentation masks with audio from VGGSound.
Evaluation protocol and metrics are further discussed in Appendix~\ref{sec:metricsExpe}.
Following the experimental protocol of \citep{AVATAR}, we train on the same 10{,}000 high-frame-rate videos selected from VGGSound~\citep{vggsound}; using this subset keeps our setup directly comparable to prior results on AVATAR.
We instantiate the multi-resolution AiSP module with pooling factors $K=\{2,2,6\}$. Each adaptation block uses convolutions with kernel size $3$, for a total of $\sim 35M$ added parameters (out of $1.7B$ in PE-AV). We train for 10 epochs with a learning rate of $1e{-4}$ and a batch size of 10. For the main experiments, we set $\mu=0.01$ and $\lambda=100$ to obtain losses of a similar order of magnitude. We use adapters of size 512, applied only in the MLP blocks, for a total of $7M$ parameters. For computational efficiency, we use the PE-AV-large variant that processes 16 frames per video instead of the version that uses all frames. We report incremental compute and memory costs of LAIP in Appendix~\ref{sec:computational_overhead}.

\subsection{Quantitative results}
\paragraph{AVATAR Benchmark}
\tightpar Table~\ref{tab:mainAvatar} reports results on all four AVATAR scenarios. LAIP substantially outperforms static sound-source localization baselines that rely on a single frame, as well as TAVLO, the strongest temporal baseline in Table~\ref{tab:mainAvatar}. 
Notably, despite TAVLO operating at a substantially higher frame rate, LAIP remains clearly superior on the \textit{multi-entity} scenario, which demands the most temporal reasoning.
To isolate LAIP from the strength of PE-AV pretraining, we reimplement both EZ-VSL, FNAC and TAVLO on the same frozen PE-AV intermediate features, with comparable trainable capacity and optimization budget. Both remain far below LAIP, indicating that access to the same strong pretrained representation is insufficient to explain our gains. 
Further details on these experiments are provided in Appendix~\ref{sec:controlled_experiments}.
LAIP obtains a lower TN than the best baselines in the \textit{off-screen} scenario. As noted in~\cite{AVATAR}, however, the TN metric used for this setting suffers from a known bias. Specifically, the threshold underlying the TN metric is computed over predictions from the entire test set across all evaluation settings, by selecting the top 10\% of heatmap activations. Consequently, methods with high activations in the other three settings may result in a threshold that artificially reduces the measured false-positive rate on off-screen examples.
Overall, these results establish a new state-of-the-art range on AVATAR.
We attribute this improvement to LAIP’s ability to leverage PE-AV pretraining, which equips the model with strong audio-visual semantic representations prior to localization-specific pattern extraction.
Applying LAIP to intermediate ImageBind or LanguageBind features also consistently outperforms TAVLO across the three localization scenarios, demonstrating that the proposed pooling mechanism is not specific to PE-AV. PE-AV outperforms all other retrieval models largely because it includes a pretrained temporal encoder, whereas ImageBind and LanguageBind require an additional video encoder trained from scratch, as they lack temporal modeling. However, when the PE-AV temporal encoder is likewise trained from scratch, its performance drops to a similar level as ImageBind and LanguageBind (see Appendix~\ref{sec:metricsExpe}).
The advantage of intermediate representations further holds for ImageBind, where LAIP substantially outperforms the same model applied to final-layer features. Together with the PE-AV layer-wise analysis (see Appendix~\ref{sec:controlled_experiments}), this suggests that the spatial–semantic trade-off exploited by LAIP generalizes across retrieval backbones. Complementary results and additional implementation details are provided in Appendix~\ref{sec:controlled_experiments}.
These findings highlight the utility of LAIP to unlock fine-grained spatial grounding from pretrained retrieval representations. Notably, although PE-AV was not designed for spatial localization, LAIP transforms its learned audio-visual semantics into spatially grounded predictions, substantially extending the capabilities of the original representation.

\begin{table*}[t]
\centering
\caption{Segmentation performance comparison across AVATAR scenarios.}
\label{tab:mainAvatar}
\small
\setlength{\tabcolsep}{5pt}
\renewcommand{\arraystretch}{1.0}

\resizebox{\textwidth}{!}{%
\begin{tabular}{l c@{\hspace{2pt}}c c@{\hspace{2pt}}c c@{\hspace{2pt}}c c}
\toprule
& \multicolumn{2}{c}{(1) Single-sound} 
& \multicolumn{2}{c}{(2) Mixed-sound} 
& \multicolumn{2}{c}{(3) Multi-entity} 
& (4) Off-screen \\
\cmidrule(lr){2-3} \cmidrule(lr){4-5} \cmidrule(lr){6-7} \cmidrule(lr){8-8}
\textbf{Method} 
& \shortstack{\textbf{CIoU(\%)}\\$\uparrow$} 
& \shortstack{\textbf{AUC(\%)}\\$\uparrow$}
& \shortstack{\textbf{CIoU(\%)}\\$\uparrow$} 
& \shortstack{\textbf{AUC(\%)}\\$\uparrow$}
& \shortstack{\textbf{CIoU(\%)}\\$\uparrow$} 
& \shortstack{\textbf{AUC(\%)}\\$\uparrow$}
& \shortstack{\textbf{TN(\%)}\\$\uparrow$} \\
\midrule
EZ-VSL(10k) \citep{EZ-VSL}      &  9.66 & 11.07 &  8.16 &  9.35 &  6.87 &  8.32 & \textbf{96.91} \\
EZ-VSL(144k) \citep{EZ-VSL}     & 10.92 & 12.22 &  6.97 &  8.34 &  5.80 &  7.42 & 96.47 \\
FNAC (144k) \citep{fnac} & 11.17 & 13.21 & 5.87 & 7.73 & 3.77 & 5.79 & 92.84 \\
FNAC (10k) \citep{fnac} & 13.00 & 14.19 & 10.99 & 12.08 & 10.31 & 11.45 & 87.52 \\
TAVLO(10k) \citep{AVATAR}   & 13.42 & 14.08 & 14.13 & 14.52 & 12.08 & 12.69 & 91.18 \\
\midrule
EZ-VSL on PE features (10k) \citep{EZ-VSL}   & 13.08 & 13.97 & 12.12 & 12.78 & 11.10 & 11.91 & 92.62 \\
FNAC on PE features (10k) \citep{fnac} & 13.55 & 13.97 & 10.98 & 11.25 & 8.89 & 9.42 & 93.47 \\
TAVLO on PE features (10k) \citep{AVATAR}  & 7.51 & 8.01 & 7.50 & 7.75 & 5.78 & 6.18 & 91.33 \\
\midrule
LAIP on ImageBind (10k) \citep{imagebind} & 18.29 & 18.56 & 16.14 & 16.30 & 15.60 & 16.02 & 90.05 \\
LAIP on LanguageBind (10k) \citep{languagebind} & 17.64 & 18.08 & 15.73 & 15.91 & 14.22 & 14.74 & 91.00 \\
\rowcolor{gray!12} LAIP (Ours, 10k) & \textbf{27.63} & \textbf{27.77} & \textbf{27.35} & \textbf{27.40} & \textbf{23.69} & \textbf{23.85} & 90.94 \\
\bottomrule
\end{tabular}%
}
\end{table*}
\paragraph{AVSBench and ADE20k} \tightpar Tables~\ref{tab:AVSBench} and~\ref{tab:SoundPrompted} show that the same model also transfers well to other sound source localization benchmarks, even though LAIP is only trained on a very small subset of VGGSound ($\sim$10k videos). On AVSBench, LAIP achieves the best F-score on both S4 and MS3 by a large margin, and also obtains the best mask-IoU on the more challenging MS3 split. The gap between F-score and mask-IoU is informative. In AVSBench, the F-score is computed with an adaptive threshold, whereas mask-IoU uses a fixed threshold of $0.5$. 
Because our method produces attention maps rather than dense segmentation logits, its predictions are naturally sharp and peaky: this structure is then penalized by having an arbitrarly fixed threshold at 0.5. This is also why LAIP is especially strong in F-score while remaining only competitive on S4 mask-IoU. For all AVSBench results, we apply min--max normalization to rescale each predicted map to $[0,1]$ before evaluation.

\begin{table*}[ht]
\centering
\caption{Quantitative results on the AVSBench test sets.}
\label{tab:AVSBench}
\begin{tabular}{lcccc}
\toprule
\multirow{2}{*}{Method} & \multicolumn{2}{c}{S4} & \multicolumn{2}{c}{MS3} \\
 & mask-IoU $\uparrow$ & F-score $\uparrow$ & mask-IoU $\uparrow$ & F-score $\uparrow$ \\
\midrule
SLAVC \citep{CloserLook}  & 28.10 & 34.60 & 24.37 & 25.56  \\
MarginNCE \citep{marginnce} & 33.27 & 45.33 & 27.31 & 31.56  \\
FNAC \citep{fnac} & 27.15 & 31.40 & 21.98 & 22.50  \\
Alignment \citep{CrossModalAlignment} & 29.60 & 35.90 & - & - \\
TACO \citep{taco} & 29.68 & 41.91 & 25.88 & 30.72 \\
\rowcolor{gray!12} LAIP (Ours) & \textbf{39.31} & \textbf{65.18} & \textbf{30.77} & \textbf{49.00} \\
\bottomrule
\end{tabular}
\end{table*}

On ADE-SP, LAIP outperforms all baselines in both m-IoU and mAP, reaching 33.35 m-IoU and 53.57 mAP. These gains are particularly notable because several competing methods are explicitly designed for sound-prompted segmentation, whereas our model is primarily built to recover localization from a retrieval-trained backbone. We believe this transfer performance stems from the same property observed on AVATAR: the pretrained PE-AV features provide strong semantic alignment, while AiSP extracts spatial evidence without destroying the retrieval representation. 

\begin{wraptable}{r}{0.48\textwidth}
    \centering
    \caption{Quantitative results on the ADE-SP dataset.}
    \label{tab:SoundPrompted}
    \resizebox{0.48\textwidth}{!}{
    \begin{tabular}{lccc}
    \toprule
        \textbf{Method} & \textbf{Retrieval model} & \textbf{m-IoU}$\uparrow$ & \textbf{mAP}$\uparrow$ \\
        \midrule
        DAVENet \citep{davenet} & \xmark & 17.0  & 16.8 \\
        DenseAV \citep{denseAV} & \xmark & 25.5  & 32.4 \\
        TACO \citep{taco} & \xmark & 27.74  & 35.75 \\
        \midrule
        ImageBind \citep{imagebind} & \cmark & 18.3  & 18.1 \\
        CAVMAE \citep{cavmae} & \cmark & 20.6  & 21.2 \\
        CAV-MAE Sync \citep{cavmaeSync} & \cmark & 22.7 & 22.6 \\
        \rowcolor{gray!12} LAIP (Ours) & \cmark & \textbf{33.35} & \textbf{53.57} \\
    \bottomrule
    \end{tabular}}
\end{wraptable}

Interestingly, our method also outperforms methods like TACO (reported without the segmenter for fair comparison), and DenseAV, which also rely on pretrained models and explicitly target sound source localization, while leaving retrieval aside. This highlights the effectiveness of our method and suggests that using an aligned audio-visual backbone is better than using separate audio and image backbones.\\
\tightpar Finally, CAV-MAE Sync~\citep{cavmaeSync} is the closest work to ours in terms of applications, since it also aims to support both sound source localization and retrieval. However, it was trained from scratch on a limited amount of data compared to PE-AV, while our method preserves compatibility with the pretrained stack and therefore benefits from higher-quality features. This difference is reflected in the results: LAIP substantially outperforms CAV-MAE Sync on ADE-SP.

\begin{table*}[t]
\centering
\normalsize
\caption{AVATAR ablations (averaged over scenarios): regularization (left) and baselines (right).}
\label{tab:ablations}

\begin{minipage}{0.48\linewidth}
\centering
\setlength{\tabcolsep}{4pt}
\begin{tabular}{lcc}
\toprule
Method & CIoU(\%) $\uparrow$ & AUC(\%) $\uparrow$ \\
\midrule
\rowcolor{gray!12}
LAIP (Ours) & \textbf{26.22} & \textbf{26.34} \\
$\mu=0$ & 23.01 & 23.27 \\
No null token & 23.11 & 23.25 \\
$\mu=\lambda=0$ & 19.45 & 19.53 \\
$\lambda=0$ & 17.52 & 17.57 \\
\bottomrule
\end{tabular}
\end{minipage}
\hfill
\begin{minipage}{0.48\linewidth}
\centering
\setlength{\tabcolsep}{4pt}
\begin{tabular}{lcc}
\toprule
Method & CIoU(\%) $\uparrow$ & AUC(\%) $\uparrow$ \\
\midrule
\rowcolor{gray!12}
LAIP (Ours) & \textbf{26.22} & \textbf{26.34} \\
Gradient-based & 4.14 & 4.89 \\
Attention pooling & 14.34 & 14.74 \\
LAIP on last layer & 15.92 & 16.38 \\
Transformer LAIP & 21.61 & 21.75 \\
\bottomrule
\end{tabular}
\end{minipage}

\end{table*}
\paragraph{Ablations} \tightpar Table~\ref{tab:ablations} shows the performance of different ablation on the AVATAR benchmark (the average scores over the scenarios are reported, but the full Tables are available in Appendix).
Removing the null-token regularization ($\mu=0$, while retaining the null token) degrades performance from $26.22/26.34$ to $23.01/23.27$ CIoU/AUC. Removing the null token entirely gives a similar $23.11/23.25$, showing that the token alone provides no benefit without the anti-collapse regularizer. Removing both $\mu$ and the multi-resolution consistency term $\lambda$ leads to a larger drop to $19.45/19.53$. These results show that preventing the attention to null token to collapse and enforcing cross-scale consistency are both important. Additional sensitivity experiments on the loss weights, number of heads, adapter size, and pooling schedule are reported in Appendix~\ref{sec:hp_sensitivity}. Additionally, replacing the CNN layers with transformer encoder layers in the AiSP modules also hurts performance ($21.61/21.75$), suggesting that local convolutions are better suited to progressive spatial refinement and alignment (as observed in~\cite{fcclip}).
Table~\ref{tab:ablations} (right) further shows that the gains cannot be recovered through simpler ways of extracting localization from PE-AV. Gradient attribution based on visual-token norms performs poorly (4.14/4.89). Using a single attention-pooling layer after the same CNN blocks is also substantially weaker (14.34/14.74 versus 26.22/26.34), showing that cross-attention alone is insufficient. Finally, applying LAIP to the final PE-AV layer degrades performance to 15.92/16.38; the complete layer-wise analysis in Appendix C confirms a clear intermediate-layer optimum. These experiments validate empirically that LAIP's performance stems from progressively extracting audio-conditioned spatial information from the appropriate intermediate representation, not merely from layer selection, cross-attention, or strong PE-AV features.

\subsection{Qualitative analysis}
\begin{figure}[t]
\centering

\newcommand{\sqimg}[1]{%
\includegraphics[width=0.085\textwidth,height=0.085\textwidth,keepaspectratio=false]{#1}
}

\setlength{\tabcolsep}{13pt} 

\begin{tabular}{cccccc}
\textbf{Input} & \textbf{TACO} & \textbf{LAIP} & \textbf{Input} & \textbf{TACO} & \textbf{LAIP} \\

\sqimg{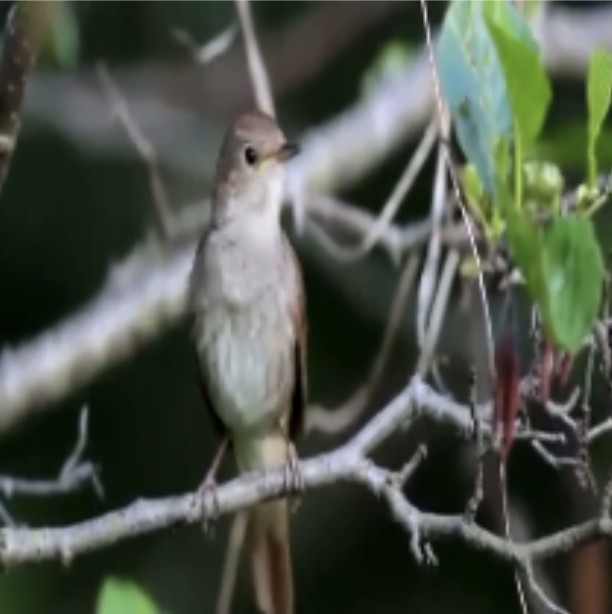} &
\sqimg{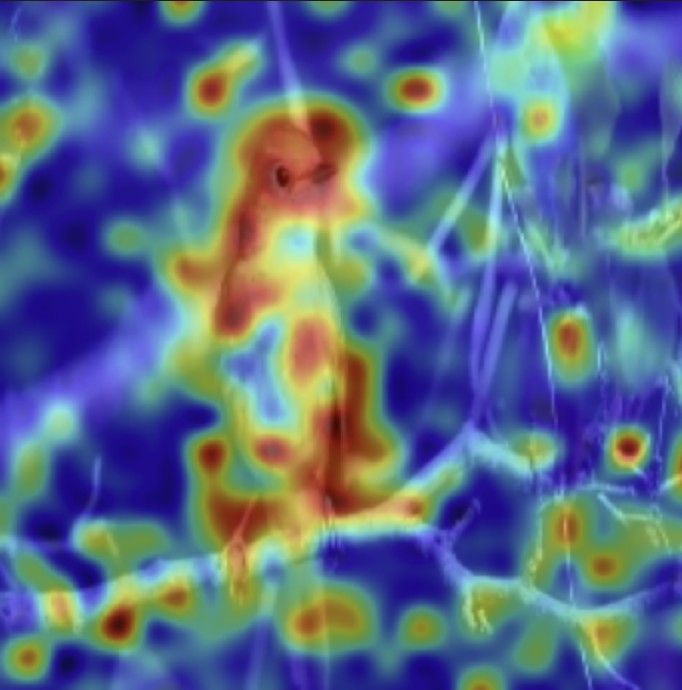} &
\sqimg{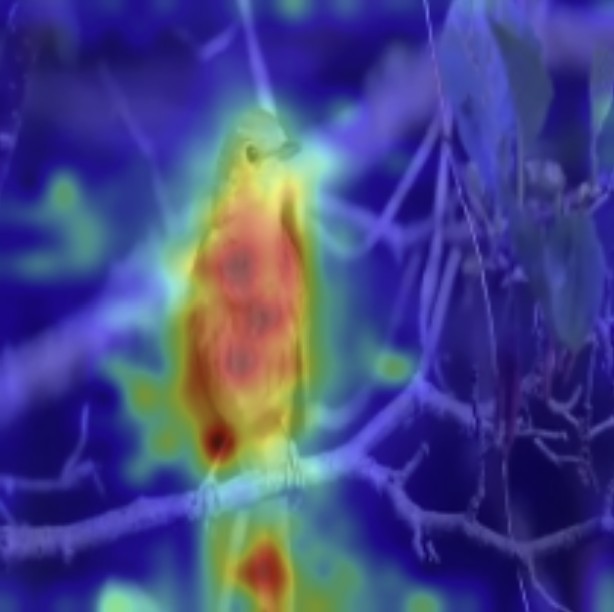} &
\sqimg{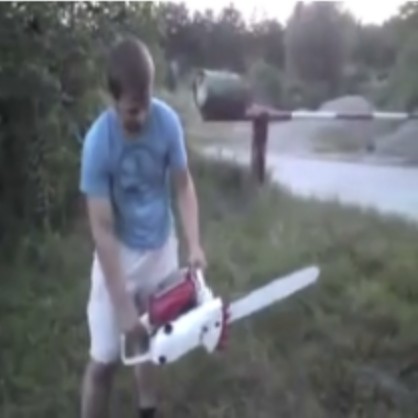} &
\sqimg{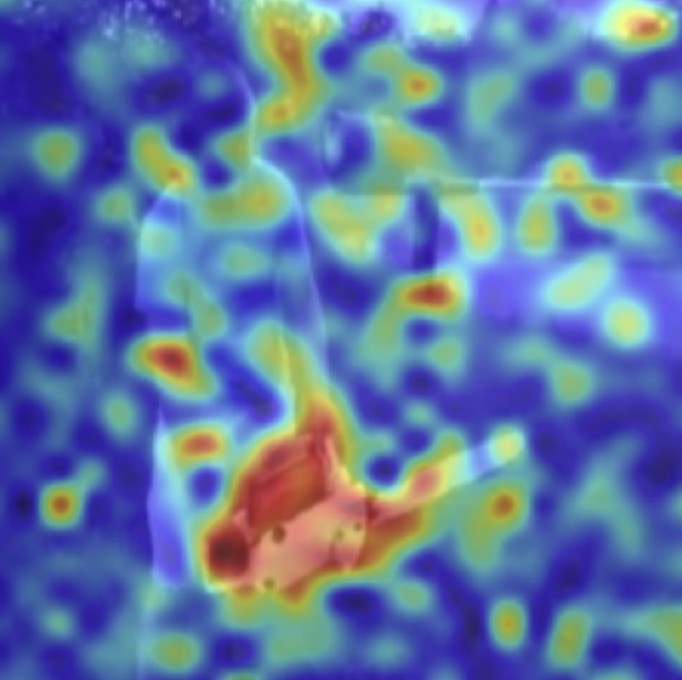} &
\sqimg{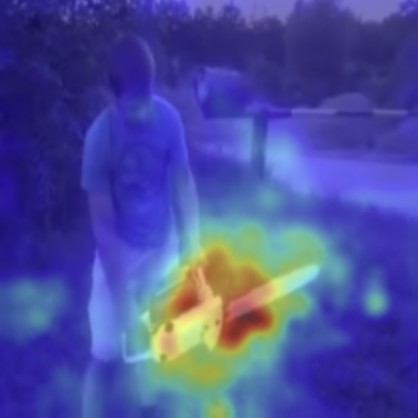} \\

\sqimg{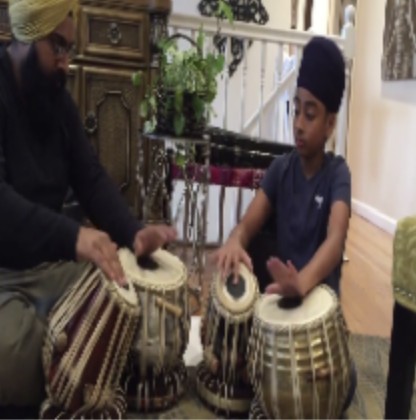} &
\sqimg{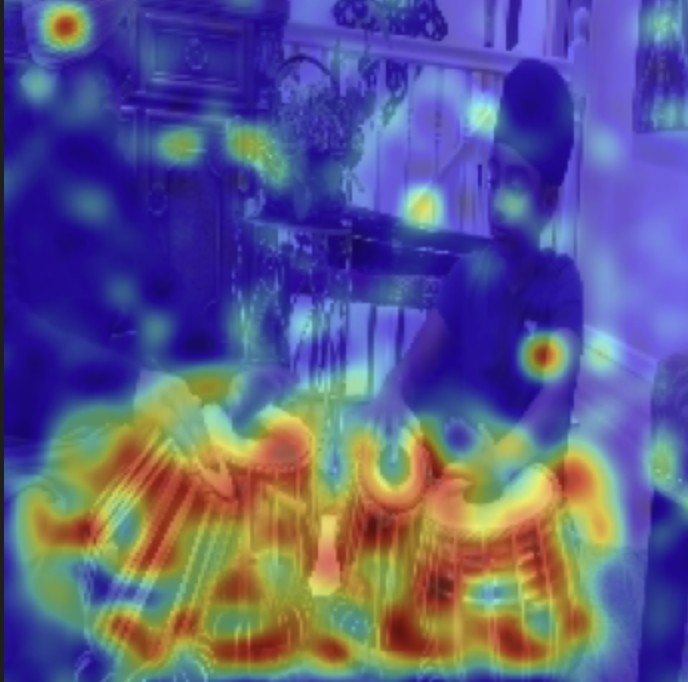} &
\sqimg{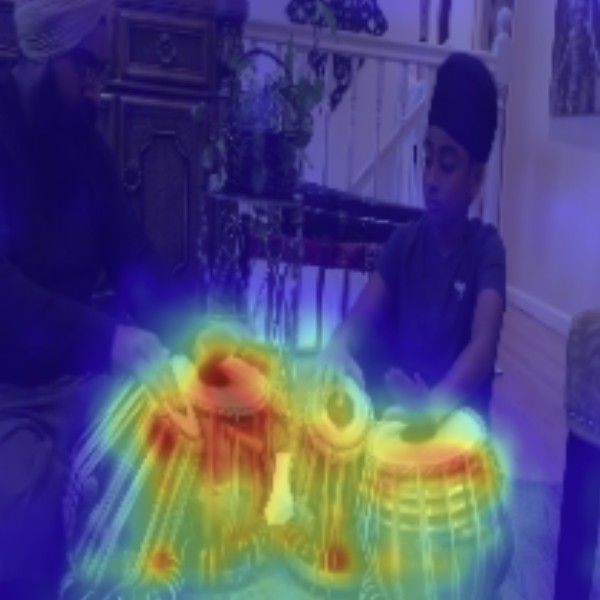} &
\sqimg{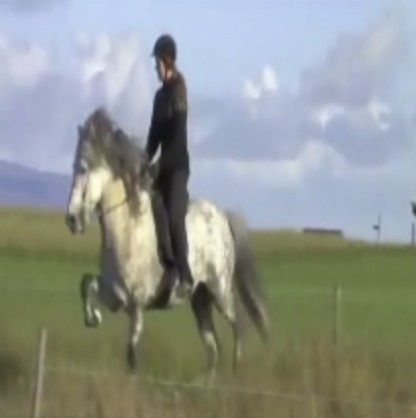} &
\sqimg{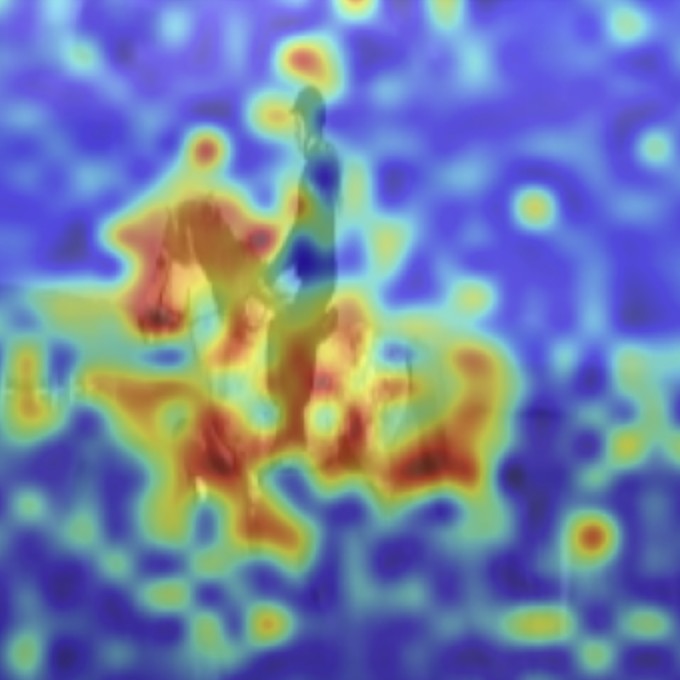} &
\sqimg{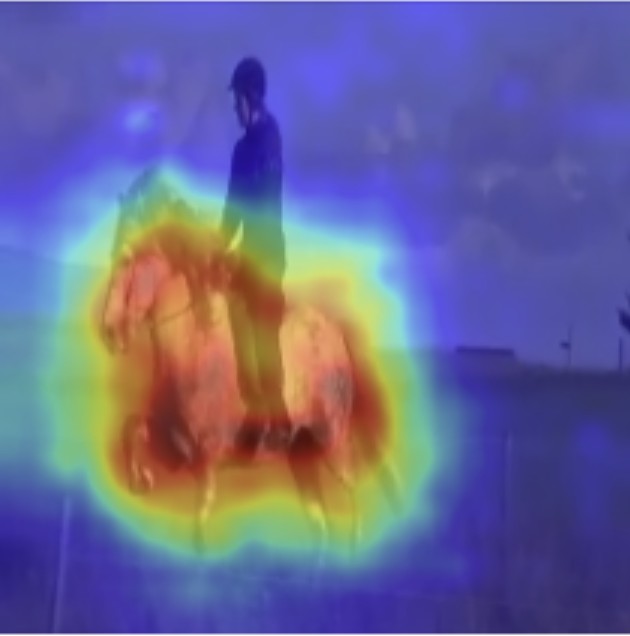} \\
\end{tabular}

\caption{TACO–LAIP comparison on S4. LAIP produces smoother maps with fewer outliers.}
\label{fig:qualiVsTACO}

\end{figure}

\tightpar Figure~\ref{fig:qualiVsTACO} compares the non-thresholded localization maps produced by TACO\cite{taco} (the best competitor both in Table~\ref{tab:AVSBench} and Table~\ref{tab:SoundPrompted}) and LAIP. This comparison is particularly informative because both methods build on large-scale pretrained models---OpenCLIP\cite{openCLIP} and CLAP\cite{CLAP} for TACO, and PE-AV\cite{peav} for LAIP---but they exploit them in different ways. TACO discovers correspondences between independently pretrained backbones through inference-time optimization, whereas LAIP learns an explicit localizer module during training. Although both methods usually identify the correct sound source, the maps produced by TACO contain many outliers, likely because of irregularities in the image-token representation. In contrast, LAIP produces smoother and more spatially coherent maps thanks to its multi-resolution pooling mechanism.

\section{Conclusion and limitations}
\tightpar We adapt a large-scale audio–visual retrieval model for sound source localization without dense spatial supervision. LAIP uses lightweight Audio-informed Spatial Pooling (AiSP) modules to recover spatial information from intermediate visual representations while preserving compatibility with the pretrained retrieval pipeline. Applied to PE-AV, this yields strong results across AVATAR, AVSBench, and ADE-SP, while experiments with ImageBind and LanguageBind further show that the proposed pooling mechanism is not specific to a single backbone.
Overall, our results suggest that global retrieval representations need not be discarded for localization: spatial information can be recovered before global aggregation and exploited within the pretrained architecture.
\\
\tightpar LAIP's main limitation is retrieval scalability: audio-conditioned pooling prevents precomputed video embeddings. Standard retrieval remains available through the unchanged PE-AV backbone; future work should enable query-independent LAIP representations.

\clearpage


\newpage
\bibliographystyle{iclr2027_conference}
\bibliography{references}

\newpage
\appendix
\section{Multi-resolution analysis}
Figure~\ref{fig:qualiStages} shows the attention maps extracted in each AiSP module in our model, as well as the aggregated prediction and the segmentation obtained by CAV-MAE Sync in the same example (from the S4 dataset). This example is particularly interesting as both models were trained (or pretrained) using at least one global objective
Interestingly, the latter focuses on the sky because helicopter sounds often co-occur with sky regions in the training data, which encourages this broad contextual association.
On the other hand the first segmentation obtained by our module produces a very accurate segmentation of the helicopter but also focuses on the sky. However, as the resolution downsizes the segmentation obtained focuses more and more on the helicopter. Hence the final aggregated segmentation is accurate and focuses much more on the helicopter than the sky. The example clearly illustrates the advantage of our multi-resolution AiSP module.
Table~\ref{tab:stages} shows the performance of the different resolutions attention maps. While combining them always improves performance, the low-resolution one is particularly accurate in the single-source setting.

\begin{minipage}{0.48\textwidth}
\centering
\small
\setlength{\tabcolsep}{2pt}
\begin{tabular}{ccc}
\includegraphics[width=0.25\linewidth,height=0.25\linewidth]{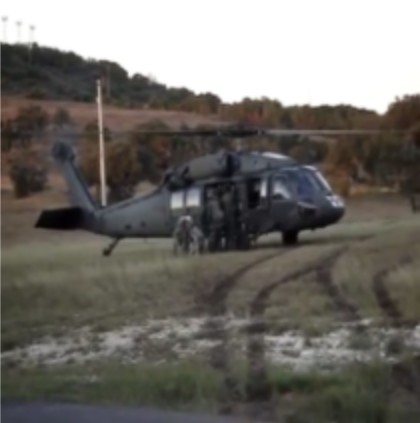} &
\includegraphics[width=0.25\linewidth,height=0.25\linewidth]{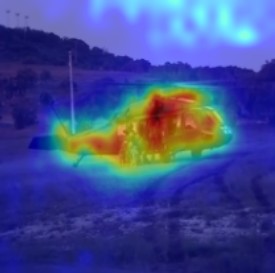} &
\includegraphics[width=0.25\linewidth,height=0.25\linewidth]{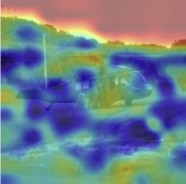} \\

Input & Average map & CMAE-Sync \\[3pt]

\includegraphics[width=0.25\linewidth,height=0.25\linewidth]{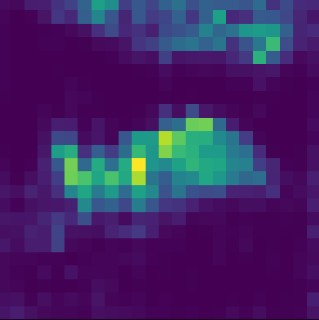} &
\includegraphics[width=0.25\linewidth,height=0.25\linewidth]{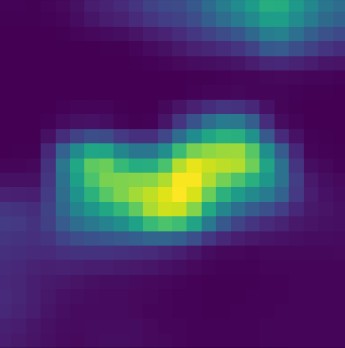} &
\includegraphics[width=0.25\linewidth,height=0.25\linewidth]{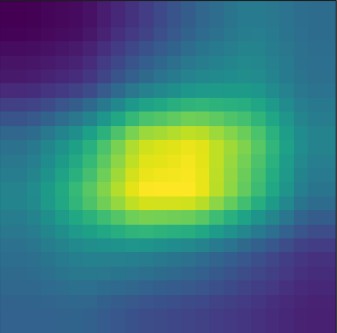} \\

First map & Second map & Third map
\end{tabular}
\captionsetup{hypcap=false}
\captionof{figure}{Qualitative example on S4.}
\label{fig:qualiStages}
\end{minipage}
\hfill
\begin{minipage}{0.48\textwidth}
\centering
\normalsize
\captionsetup{hypcap=false}
\captionof{table}{Performance by stage on the S4 and MS3 subsets. Averaging the stage-wise maps gives the best results on both datasets.}
\label{tab:stages}
\setlength{\tabcolsep}{4pt}
\renewcommand{\arraystretch}{1.1}
\begin{tabular}{lcc}
\toprule
Method & S4 mIoU $\uparrow$ & MS3 mIoU $\uparrow$ \\
\midrule
Low & 36.60 & 27.10 \\
Mid & 34.57 & 28.58 \\
High & 29.24 & 27.43 \\
\midrule
\rowcolor{gray!12}Average & \textbf{39.31} & \textbf{30.77} \\
\bottomrule
\end{tabular}
\end{minipage}

\section{Additional metrics and experiments}
\label{sec:metricsExpe}
In this section, we report detailed metrics for the ablations, as the main paper only shows the average across scenarios. Additionally, we also detail the meaning of the metrics and report additional experiments. 

On AVATAR, CIoU measures the overlap between the thresholded heatmap and the annotated sounding region, AUC summarizes localization quality across thresholds, and TN is used for off-screen videos where the correct behavior is to suppress activation. On AVSBench, we report mask-IoU (noted as mIoU in some tables) and F-score: mask-IoU measures the overlap between the predicted binary sounding mask and the ground-truth mask using the benchmark's fixed threshold, while F-score uses an adaptive threshold and therefore better reflects whether the method highlights the correct sounding region even when the predicted map is sharp or imperfectly calibrated. On ADE-SP, we report maskIoU as well as mAP: whereas mAP summarizes precision--recall performance over the predicted maps and is therefore sensitive to ranking quality beyond a single threshold. Higher is better for all metrics.

For simplicity, we only comment the ablations that were not included in the main paper.

\begin{table*}[ht]
\centering
\caption{Ablation study on AVATAR across the four evaluation scenarios.}
\label{tab:ablation_scenarios}
\small
\setlength{\tabcolsep}{5pt}
\renewcommand{\arraystretch}{1.1}

\resizebox{\textwidth}{!}{%
\begin{tabular}{lccccccc}
\toprule
& \multicolumn{2}{c}{(1) Single-sound} 
& \multicolumn{2}{c}{(2) Mixed-sound} 
& \multicolumn{2}{c}{(3) Multi-entity} 
& (4) Off-screen \\
\cmidrule(lr){2-3} \cmidrule(lr){4-5} \cmidrule(lr){6-7} \cmidrule(lr){8-8}
\textbf{Method} 
& \shortstack{\textbf{CIoU(\%)}\\$\uparrow$} 
& \shortstack{\textbf{AUC(\%)}\\$\uparrow$}
& \shortstack{\textbf{CIoU(\%)}\\$\uparrow$} 
& \shortstack{\textbf{AUC(\%)}\\$\uparrow$}
& \shortstack{\textbf{CIoU(\%)}\\$\uparrow$} 
& \shortstack{\textbf{AUC(\%)}\\$\uparrow$}
& \shortstack{\textbf{TN(\%)}\\$\uparrow$} \\
\midrule
\rowcolor{gray!12} LAIP (Ours) & \textbf{27.63} & \textbf{27.77} & \textbf{27.35} & \textbf{27.40} & \textbf{23.69} & \textbf{23.85} & 90.94 \\
$\mu=0$& 24.46 & 24.75 & 24.60 & 24.72 & 19.97 & 20.34 & \textbf{92.45} \\
No null token & 24.37 & 24.55 & 24.18 & 24.20 & 20.79 & 21.01 & 90.89 \\
$\mu=\lambda=0$& 19.59 & 19.70 & 21.18 & 21.18 & 17.57 & 17.71 & 90.76 \\
$\lambda=0$ & 17.13 & 17.17  & 19.82 & 19.85 & 15.61 & 15.69 & 89.35 \\
Transformer LAIP & 22.01 & 22.14 & 23.43 & 23.51 & 19.40 & 19.59 & 90.77 \\
Attention pooling (1 head) & 15.09 & 15.59 & 13.70 & 14.02 & 14.22 & 14.62 & 91.31 \\
AiSP with a single final head & 22.07 & 22.36 & 22.60 & 22.74 & 18.34 & 18.74 & 91.76 \\
Attention pooling (avg heads) & 20.88 & 21.18 & 23.89 & 24.02 & 17.89 & 18.15 & 92.21 \\

\bottomrule
\end{tabular}%
}

\end{table*}

Removing the null token yields $23.11/23.25$ CIoU/AUC on the AVATAR (averaged across the three localization scenarios), nearly matching the $\mu=0$ variant in which the token remains present but unregularized. This indicates that simply adding the null token is insufficient; its benefit emerges when the null-attention penalty prevents it from becoming a shortcut on negative pairs. We analyze this behavior more in depth in Appendix \ref{sec:hardNeg}.

In the main paper, we ablate the AiSP module by replacing it with a CNN followed by a single attention-pooling layer. Since the multi-resolution loss is removed in this setting, the attention head can no longer be selected automatically; therefore, the experiment was conducted using a single attention head.
\\
To ensure a fair comparison, we also evaluated our localization module using a single head in the final AiSP layer. Although performance decreases compared to the model with 8 attention heads, it remains substantially better than the attention-pooling baseline using the same number of heads.
\\
Conversely, we evaluated the attention-pooling baseline with the same number of heads as our main model (8) and averaged the outputs at inference time. While this improves performance over the single-head version, it still remains significantly below LAIP with multiple attention heads.

\begin{table*}[ht]
\centering
\caption{Ablation study on AVATAR across the four evaluation scenarios.}
\label{tab:ablation_video_full}
\small
\setlength{\tabcolsep}{5pt}
\renewcommand{\arraystretch}{1.1}

\resizebox{\textwidth}{!}{%
\begin{tabular}{lccccccc}
\toprule
& \multicolumn{2}{c}{(1) Single-sound} 
& \multicolumn{2}{c}{(2) Mixed-sound} 
& \multicolumn{2}{c}{(3) Multi-entity} 
& (4) Off-screen \\
\cmidrule(lr){2-3} \cmidrule(lr){4-5} \cmidrule(lr){6-7} \cmidrule(lr){8-8}
\textbf{Method} 
& \shortstack{\textbf{CIoU(\%)}\\$\uparrow$} 
& \shortstack{\textbf{AUC(\%)}\\$\uparrow$}
& \shortstack{\textbf{CIoU(\%)}\\$\uparrow$} 
& \shortstack{\textbf{AUC(\%)}\\$\uparrow$}
& \shortstack{\textbf{CIoU(\%)}\\$\uparrow$} 
& \shortstack{\textbf{AUC(\%)}\\$\uparrow$}
& \shortstack{\textbf{TN(\%)}\\$\uparrow$} \\
\midrule
\rowcolor{gray!12} LAIP (Ours) & \textbf{27.63} & \textbf{27.77} & \textbf{27.35} & \textbf{27.40} & \textbf{23.69} & \textbf{23.85} & 90.94 \\
LAIP with global audio query &
26.40 & 26.52 &
27.91 & 27.97 &
22.57 & 22.69 &
91.27 \\
Gradient-based & 4.35 & 5.06 & 3.49 & 4.39 & 4.58 & 5.22 & 88.17 \\
LAIP on last PE layer & 16.45 & 17.02 & 17.17 & 17.42 & 14.14 & 14.70 & 89.76 \\
LAIP video encoder scratch & 19.03 & 19.35 & 22.80 & 22.88 & 17.59 & 17.82 & 92.03 \\
LAIP no adapter & 20.47 & 20.54 & 18.96 & 19.02 & 17.20 & 17.37 & 90.58 \\
\bottomrule
\end{tabular}%
}

\end{table*}
To assess the role of temporal audio alignment, we replace the frame-specific audio query at each time step with a single globally aggregated audio representation shared across all frames. This variant obtains $25.45/25.63$ CIoU/AUC when averaging the three AVATAR scenarios, compared with $26.22/26.34$ for frame-aligned LAIP. Frame-aligned queries improve Single-Sound and Multi-Entity localization, whereas the global query is slightly better on Mixed-Sound and off-screen rejection. Thus, local audio-frame alignment provides a modest overall gain, while global audio context remains competitive. We hypothesize that the relatively small gap results from temporal over-smoothing in the CNN layers, which reduces the temporal specificity of the audio tokens and, consequently, their alignment with individual frames. Explicitly encouraging temporally distinct audio representations may alleviate this effect; we leave this direction for future work.

The rows added in Table~\ref{tab:ablation_video_full} but omitted from the main paper further clarify the role of the pretrained temporal stack. Training the video encoder from scratch performs substantially worse than full LAIP on the three localization scenarios, showing that a part of the gain comes from preserving the pretrained video encoder rather than relearning temporal aggregation from the limited AVATAR training set. Similarly, removing the adapters also degrades performance, which confirms that these lightweight modules are important to bridge our audio-informed token with the pretrained video encoder without collapsing the representation to an estimation of the original [CLS] token.

\begin{table*}[ht]
\centering
\caption{Ablation study on AVATAR across the four evaluation scenarios.}
\label{tab:aisp_number}
\small
\setlength{\tabcolsep}{5pt}
\renewcommand{\arraystretch}{1.1}

\resizebox{\textwidth}{!}{%
\begin{tabular}{lccccccc}
\toprule
& \multicolumn{2}{c}{(1) Single-sound} 
& \multicolumn{2}{c}{(2) Mixed-sound} 
& \multicolumn{2}{c}{(3) Multi-entity} 
& (4) Off-screen \\
\cmidrule(lr){2-3} \cmidrule(lr){4-5} \cmidrule(lr){6-7} \cmidrule(lr){8-8}
\textbf{Method} 
& \shortstack{\textbf{CIoU(\%)}\\$\uparrow$} 
& \shortstack{\textbf{AUC(\%)}\\$\uparrow$}
& \shortstack{\textbf{CIoU(\%)}\\$\uparrow$} 
& \shortstack{\textbf{AUC(\%)}\\$\uparrow$}
& \shortstack{\textbf{CIoU(\%)}\\$\uparrow$} 
& \shortstack{\textbf{AUC(\%)}\\$\uparrow$}
& \shortstack{\textbf{TN(\%)}\\$\uparrow$} \\
\midrule
\rowcolor{gray!12} Ours & \textbf{27.63} & \textbf{27.77} & \textbf{27.35} & \textbf{27.40} & \textbf{23.69} & \textbf{23.85} & 90.94 \\
LAIP Pooler inverted bottlneck (4 3 2) & 13.67 & 13.94 & 15.01 & 15.12 & 11.95 & 12.38 & 90.83 \\
LAIP two poolers &21.33&21.45& 23.21&23.26&17.51&17.70 &90.52 \\
\bottomrule
\end{tabular}%
}
\end{table*}
Table~\ref{tab:aisp_number} comment on the pooling design itself. ``LAIP two poolers'' means that we use only two AiSP modules instead of the default three-stage hierarchy. This variant remains clearly below full LAIP, which suggests that three successive pooling stages are useful to progressively refine the audio-conditioned spatial evidence before producing a token compatible with the video encoder. ``LAIP Pooler inverted bottleneck (4 3 2)'' reverses our usual compression schedule by applying the strongest spatial reduction first and smaller reductions later. This setting performs much worse on all three localization scenarios, indicating that aggressive early compression destroys fine spatial cues before the later AiSP modules can exploit them. Together, these ablations support our default design choice: gradual compression with three successive AiSP modules.

\begin{figure*}[t]
    \centering

    \begin{subfigure}{0.24\linewidth}
        \centering
        \includegraphics[width=\linewidth]{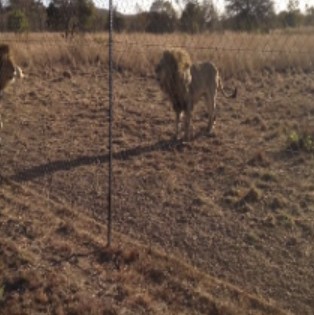}
        \caption{Input image}
        \label{fig:sub1}
    \end{subfigure}
    \hfill
    \begin{subfigure}{0.24\linewidth}
        \centering
        \includegraphics[width=\linewidth]{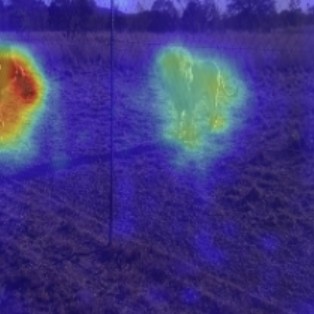}
        \caption{Heatmap}
        \label{fig:sub2}
    \end{subfigure}
    \hfill
    \begin{subfigure}{0.24\linewidth}
        \centering
        \includegraphics[width=\linewidth]{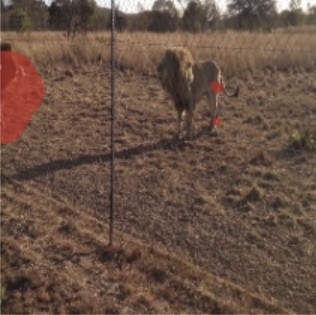}
        \caption{mIoU threshold}
        \label{fig:sub3}
    \end{subfigure}
    \hfill
    \begin{subfigure}{0.24\linewidth}
        \centering
        \includegraphics[width=\linewidth]{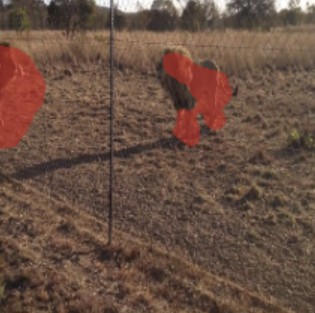}
        \caption{F-score threshold}
        \label{fig:sub4}
    \end{subfigure}

    \caption{Impact of the thresholding on attention maps.}
    \label{fig:threshold}
\end{figure*}

Finally, the main paper shows that LAIP can be much stronger in F-score than in mask-IoU on AVSBench. We argue that this is due to the fact that our method produces attention maps rather than dense segmentation logits, and these maps are often sharp and peaky around the true sounding region. As a result, after binarization with the benchmark's fixed threshold of $0.5$, part of the predicted support can be removed too aggressively, which hurts mask-IoU even when the localization itself is correct. The adaptive threshold used for F-score is more tolerant to this calibration issue, which is why it better reflects the quality of our localization maps.
Figure~\ref{fig:threshold} illustrates this behaviour: while the heatmap effectively highlights the two lions, binarizing with the 0.5 threshold excludes the second lion, whereas using the adaptive threshold in the F-score yields a much better segmentation.

\section{Analysis of the null token regularization}
\label{sec:hardNeg}
\begin{figure}
    \centering
    \includegraphics[width=0.5\linewidth]{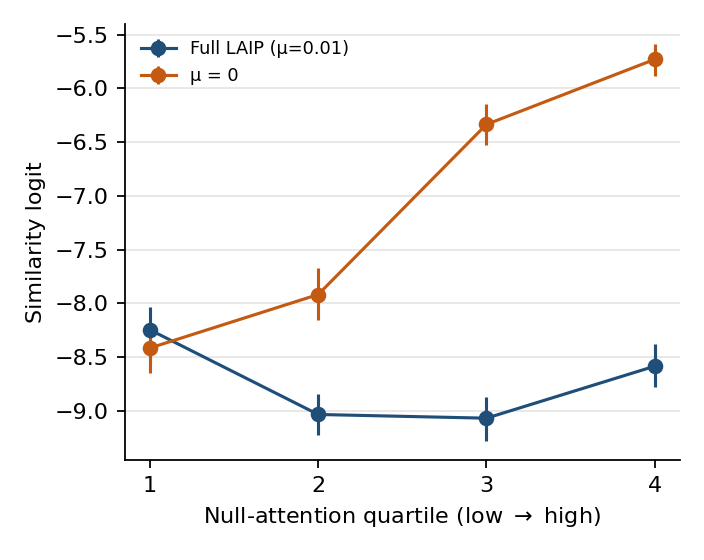}
    \caption{Audio–visual similarity of hard-negative pairs as a function of the attention assigned to the null token.}
    \label{fig:hardneg}
\end{figure}

To better understand the role of null-token regularization, we analyze hard negatives, defined as the mismatched audio-video pairs with the highest cosine similarity under the frozen PE-AV model. We group these pairs into quartiles according to the attention assigned to the null token. As shown in Figure~\ref{fig:hardneg}, without regularization ($\mu=0$), larger null attention is associated with substantially higher similarity logits for negative pairs, whereas the regularized model keeps them consistently separated. This exposes a failure mode of the unregularized null token: because it participates in the attention normalization but is excluded from the pooled visual representation, assigning excessive mass to it leaves only a weak contribution from the actual spatial tokens. For hard negatives, this weakens the visual evidence and makes the resulting representation less discriminative, bringing mismatched audio-visual pairs closer in the embedding space. Penalizing null attention prevents this behavior and preserves informative visual evidence even for difficult negative pairs.

\section{Additional details on controlled experiments}
\label{sec:controlled_experiments}
\paragraph{EZ-VSL on PE-AV features.}
Here, we detail the implementation used to train EZ-VSL on PE features.
To ensure a fair comparison with LAIP, we train a three-layer CNN on image features extracted from the 16th layer of the Perception Encoder, and a two-layer MLP on top of the audio features (the CLS token) extracted from the PE-AV audio encoder. The overall number of trainable parameters is $\sim30M$, which is similar to LAIP and ensures
fair comparison.
Since contrastive methods typically require larger batch sizes to achieve good performance, we increased the batch size to 32 and trained the model for 40 epochs in order to obtain a number of training steps comparable to that of LAIP. The same setup was used for the adaptation of FNAC. In Table~\ref{tab:mainAvatar}, ``FNAC (144k)'' denotes FNAC trained on the full VGGSound training set, comprising approximately 144{,}000 clips, whereas ``FNAC (10k)'' uses the 10{,}000-video subset used for the AVATAR comparison. ``FNAC on PE features (10k)'' is our adaptation using frozen PE-AV features and the same 10{,}000-video subset.

\paragraph{TAVLO on PE-AV features.}
We train one factorized attention block on top of frozen PE-AV features with the TAVLO objective. The PE-AV visual tokens come from a full attention layer, so each token carries global information about the frame. Under the TAVLO loss this makes the training collapse. The loss only requires one spatial location to match the audio of the clip. Hence a token that already sees the whole frame can provide that match without covering the sounding object.

We then replace the visual encoder with a standard ResNet-18, whose cells are local, and keep the PE-AV audio tokens. This removes the global shortcut, but does significantly improve the results (Table~\ref{tab:tavlo-pe-resnet}).

\begin{table*}[htbp]
    \centering
    \small
    \setlength{\tabcolsep}{5pt}
    \renewcommand{\arraystretch}{1.1}
    \begin{tabular}{lcc}
        \toprule
        Vision encoder & CIoU & AUC \\
        \midrule
        PE-AV tokens & 7.05 & 7.54 \\
        ResNet-18    & 7.78 & 9.03 \\
        \bottomrule
    \end{tabular}
    \caption{Performance of TAVLO adapted to PE tokens}
    \label{tab:tavlo-pe-resnet}
\end{table*}

\paragraph{LAIP on ImageBind.}
To assess whether LAIP can recover spatial grounding from another pretrained audio--visual backbone, we apply it to visual tokens extracted from layer 24 of the 32-layer ImageBind vision encoder~\citep{imagebind}. Since this configuration does not include the pretrained PE-AV temporal stack, we use the same randomly initialized downstream video encoder as in the ``LAIP video encoder scratch'' ablation. All other training and evaluation settings are kept unchanged (we used the same setup for LanguageBind). We additionally apply LAIP to the final ImageBind layer while retaining the same downstream encoder. Table~\ref{tab:imagebind_full} shows that the intermediate ImageBind features substantially outperform the final-layer features across all three localization scenarios, while also outperforming TAVLO as reported in Table~\ref{tab:mainAvatar}. This provides evidence that LAIP is not specific to PE-AV and that intermediate representations retain more useful spatial information across backbones.

\begin{table*}[htbp]
\centering
\caption{LAIP performance with PE-AV and ImageBind features on the four AVATAR scenarios. Both ImageBind variants use a randomly initialized downstream video encoder.}
\label{tab:imagebind_full}
\small
\setlength{\tabcolsep}{5pt}
\renewcommand{\arraystretch}{1.1}
\resizebox{\textwidth}{!}{%
\begin{tabular}{lccccccc}
\toprule
& \multicolumn{2}{c}{(1) Single-sound}
& \multicolumn{2}{c}{(2) Mixed-sound}
& \multicolumn{2}{c}{(3) Multi-entity}
& (4) Off-screen \\
\cmidrule(lr){2-3} \cmidrule(lr){4-5} \cmidrule(lr){6-7} \cmidrule(lr){8-8}
\textbf{Method}
& \textbf{CIoU(\%)} $\uparrow$ & \textbf{AUC(\%)} $\uparrow$
& \textbf{CIoU(\%)} $\uparrow$ & \textbf{AUC(\%)} $\uparrow$
& \textbf{CIoU(\%)} $\uparrow$ & \textbf{AUC(\%)} $\uparrow$
& \textbf{TN(\%)} $\uparrow$ \\
\midrule
\rowcolor{gray!12} LAIP (PE-AV L16) & \textbf{27.63} & \textbf{27.77} & \textbf{27.35} & \textbf{27.40} & \textbf{23.69} & \textbf{23.85} & 90.94 \\
LAIP (PE-AV L16, video encoder scratch) & 19.03 & 19.35 & 22.80 & 22.88 & 17.59 & 17.82 & \textbf{92.03} \\
LAIP (ImageBind L24) & 18.29 & 18.56 & 16.14 & 16.30 & 15.60 & 16.02 & 90.05 \\
LAIP (ImageBind L32) & 5.83 & 6.88 & 5.76 & 6.79 & 5.14 & 6.28 & 89.93 \\
\bottomrule
\end{tabular}%
}
\end{table*}

\paragraph{PE-AV layer selection.}
We further evaluate LAIP using visual features extracted from layers 8, 14, 16, 18, and 24 of the PE-AV frame encoder, while keeping the remainder of the architecture and training configuration fixed. Table~\ref{tab:peav_layer_full} reveals a clear middle-layer optimum: performance improves from layer 8 to layer 16 and then decreases at layers 18 and 24. This trend suggests a spatial--semantic trade-off. Earlier layers preserve fine visual structure but are less aligned with audio semantics, whereas later layers are more semantic but have progressively discarded spatial detail. Layer 16 provides the best balance across the three localization scenarios and is therefore used in all main experiments. Together with the ImageBind results above, this analysis shows that LAIP benefits from intervening before the final representation in two different pretrained audio--visual backbones.

\begin{table*}[htbp]
\centering
\caption{Layer-wise LAIP performance on the four AVATAR scenarios using PE-AV frame-encoder features.}
\label{tab:peav_layer_full}
\small
\setlength{\tabcolsep}{5pt}
\renewcommand{\arraystretch}{1.1}
\resizebox{\textwidth}{!}{%
\begin{tabular}{lccccccc}
\toprule
& \multicolumn{2}{c}{(1) Single-sound}
& \multicolumn{2}{c}{(2) Mixed-sound}
& \multicolumn{2}{c}{(3) Multi-entity}
& (4) Off-screen \\
\cmidrule(lr){2-3} \cmidrule(lr){4-5} \cmidrule(lr){6-7} \cmidrule(lr){8-8}
\textbf{PE-AV layer}
& \textbf{CIoU(\%)} $\uparrow$ & \textbf{AUC(\%)} $\uparrow$
& \textbf{CIoU(\%)} $\uparrow$ & \textbf{AUC(\%)} $\uparrow$
& \textbf{CIoU(\%)} $\uparrow$ & \textbf{AUC(\%)} $\uparrow$
& \textbf{TN(\%)} $\uparrow$ \\
\midrule
8  & 19.51 & 19.68 & 20.32 & 20.36 & 16.09 & 16.33 & 89.73 \\
14 & 25.56 & 25.62 & 22.77 & 22.79 & 20.70 & 20.84 & 90.59 \\
\rowcolor{gray!12} 16 & \textbf{27.63} & \textbf{27.77} & \textbf{27.35} & \textbf{27.40} & \textbf{23.69} & \textbf{23.85} & \textbf{90.94} \\
18 & 23.18 & 23.27 & 23.87 & 23.89 & 20.62 & 20.73 & 90.78 \\
24 & 16.45 & 17.02 & 17.17 & 17.42 & 14.14 & 14.70 & 89.76 \\
\bottomrule
\end{tabular}%
}
\end{table*}

\section{Hyperparameter sensitivity}
\label{sec:hp_sensitivity}
All hyperparameters and model checkpoints were selected using the VGGSoundSource benchmark and then kept fixed for every evaluation benchmark. To assess sensitivity around the selected configuration, we vary one design choice at a time while retaining all remaining settings. The default configuration uses $\mu=0.01$, $\lambda=100$, $8$ heads in the final AiSP stage, adapters of size $512$, and pooling factors $K=\{2,2,6\}$. Table~\ref{tab:hp_sensitivity} shows that modestly reducing $\lambda$ to $75$, reducing the number of heads, or changing the adapter size preserves similar performance. The null-token weight and pooling schedule have a larger influence, but all non-degenerate variants retain strong localization performance. Overall, the default configuration provides the most consistent results across the three localization scenarios rather than being narrowly optimized for a single setting.

\begin{table*}[htbp]
\centering
\caption{Hyperparameter sensitivity on the four AVATAR scenarios. Each row changes only the indicated setting relative to the default LAIP configuration.}
\label{tab:hp_sensitivity}
\small
\setlength{\tabcolsep}{4.5pt}
\renewcommand{\arraystretch}{1.08}
\resizebox{\textwidth}{!}{%
\begin{tabular}{lccccccc}
\toprule
& \multicolumn{2}{c}{(1) Single-sound}
& \multicolumn{2}{c}{(2) Mixed-sound}
& \multicolumn{2}{c}{(3) Multi-entity}
& (4) Off-screen \\
\cmidrule(lr){2-3} \cmidrule(lr){4-5} \cmidrule(lr){6-7} \cmidrule(lr){8-8}
\textbf{Configuration}
& \textbf{CIoU(\%)} $\uparrow$ & \textbf{AUC(\%)} $\uparrow$
& \textbf{CIoU(\%)} $\uparrow$ & \textbf{AUC(\%)} $\uparrow$
& \textbf{CIoU(\%)} $\uparrow$ & \textbf{AUC(\%)} $\uparrow$
& \textbf{TN(\%)} $\uparrow$ \\
\midrule
\rowcolor{gray!12} LAIP (default) & 27.63 & 27.77 & \textbf{27.35} & \textbf{27.40} & \textbf{23.69} & \textbf{23.85} & 90.94 \\
$\mu=0$ & 24.46 & 24.75 & 24.60 & 24.72 & 19.97 & 20.34 & \textbf{92.45} \\
$\mu=\lambda=0$ & 19.59 & 19.70 & 21.18 & 21.18 & 17.57 & 17.71 & 90.76 \\
$\lambda=0$ & 17.13 & 17.17 & 19.82 & 19.85 & 15.61 & 15.69 & 89.35 \\
$\mu=0.001$ & 24.92 & 25.10 & 24.83 & 24.87 & 20.56 & 20.83 & 91.32 \\
$\lambda=75$ & 27.23 & 27.29 & 25.92 & 25.94 & 22.78 & 22.96 & 90.86 \\
$\lambda=75,\ \mu=0.001$ & 25.49 & 25.61 & 26.76 & 26.82 & 22.58 & 22.81 & 91.26 \\
$4$ heads & \textbf{27.91} & \textbf{28.01} & 26.09 & 26.10 & 23.44 & 23.61 & 90.92 \\
Adapter size $1024$ & 25.86 & 25.97 & 26.52 & 26.55 & 22.76 & 22.88 & 91.07 \\
$K=\{2,3,4\}$ & 22.89 & 23.06 & 25.34 & 25.39 & 21.40 & 21.54 & 91.90 \\
\bottomrule
\end{tabular}%
}
\end{table*}

\clearpage

\section{Computational overhead}
\label{sec:computational_overhead}
\paragraph{Localizer overhead}
We quantify the incremental cost of LAIP relative to its PE-AV backbone using a 10-second video and its corresponding 10-second audio clip. As reported in Table~\ref{tab:computational_overhead}, PE-AV requires $6.67$ TFLOPs, while LAIP adds only $281.6$ GFLOPs ($0.282$ TFLOPs), yielding a total cost of $6.95$ TFLOPs. This corresponds to a $4.22\%$ compute overhead. Similarly, peak memory increases from $8.45$ GiB for PE-AV to $8.79$ GiB for PE-AV with LAIP, an increase of $0.34$ GiB, or approximately $4.0\%$. Thus, LAIP adds spatial grounding while largely preserving the computational profile of the pretrained backbone. 

\begin{table}[htbp]
\centering
\caption{Incremental computational overhead of LAIP for a 10-second video-audio pair.}
\label{tab:computational_overhead}
\small
\setlength{\tabcolsep}{6pt}
\renewcommand{\arraystretch}{1.1}
\begin{tabular}{lccc}
\toprule
Resource & PE-AV & PE-AV + LAIP & Overhead \\
\midrule
Compute & $6.67$ TFLOPs & $6.95$ TFLOPs & $+4.22\%$ \\
Peak memory & $8.45$ GiB & $8.79$ GiB & $+4.0\%$ \\
\bottomrule
\end{tabular}
\end{table}

\paragraph{Comparison with other SSL methods}
The figures above are measured on the experimental setting: a native PE-AV forward at $336$ px, reported in FLOPs, with one multiply-add counted as two operations. For comparison with other localizers we now instead follow the definition used by TACO, which is not the one used in our experiments. A multiply-add counts as one MAC, and every frame is $224$ px, the resolution of the ConvNeXt-Large encoder in TACO’s cost table. LAIP is measured directly under that definition, with the visual encoder stopped at layer $-8$, the representation read by the localization head.

We rescale the single-frame MAC counts reported by TACO to a $16$-frame clip: the image encoder is counted once per frame, and a clip-level audio encoder is counted once. Table~\ref{tab:macs_224} separates that encoding cost from the localizer. CAV-MAE Sync is measured end to end, and its cosine map is $0.00024\times 10^{10}$ MACs. TACO reports no MAC count for the decomposition that produces its segmentation. The paper states that this step adds about $20\%$ wall-clock time relative to prompting FC-CLIP, after about $1000$ optimization steps on one A100, and we apply that percentage to the encoder total. The FC-CLIP segmenter itself is not included. DenseAV solely relies on two encoders. Its localization is the pairwise inner product between audio and visual tokens; on the ViT-B/16 grid implied by the reported image cost, with two heads and a $10$ s HuBERT sequence, that map is about $0.12\times 10^{10}$ MACs over $16$ frames.

The results show that LAIP is more expensive than the previous methods, but the cost stays in the same order of magnitude, and the gap is in the encoders. The localization head is $6.3\times 10^{10}$ MACs, $4.2\%$ of the LAIP total, a smaller share than TACO’s estimated decomposition ($16.7\%$). The maps used by DenseAV and CAV-MAE Sync are negligible besides their encoders ($0.35\%$ and $0.0004\%$). Localization performance is nevertheless substantially higher in LAIP.

\begin{table}[htbp]
\centering
\small
\setlength{\tabcolsep}{5pt}
\renewcommand{\arraystretch}{1.1}
\begin{tabular}{lccccc}
\toprule
Method & Vision encoding & Audio encoding & Localizer & Total & Localizer/total \\
\midrule
DenseAV & $27.0$ & $6.72$ & $0.12^{\ddagger}$ & $33.9$ & $0.35\%$ \\
TACO & $55.0$ & $0.63$ & $11.1^{\dagger}$ & $66.8$ & $16.7\%$ \\
CAV-MAE Sync & $29.2$ & $30.8$ & $0.00024$ & $60.0$ & $0.0004\%$ \\
LAIP (224 px) & $91.9$ & $52.7$ & $6.3$ & $150.9$ & $4.2\%$ \\
\bottomrule
\end{tabular}
\caption{MAC breakdown ($\times 10^{10}$) of the 16-frame totals above. $^{\dagger}$Estimated decomposition. $^{\ddagger}$Estimated similarity map.}
\label{tab:macs_224}
\end{table}

\section{Gradient-based segmentation analysis}

\begin{figure}[ht]
  \centering
  \begin{subfigure}[t]{0.23\linewidth}
    \centering
    \includegraphics[width=\linewidth]{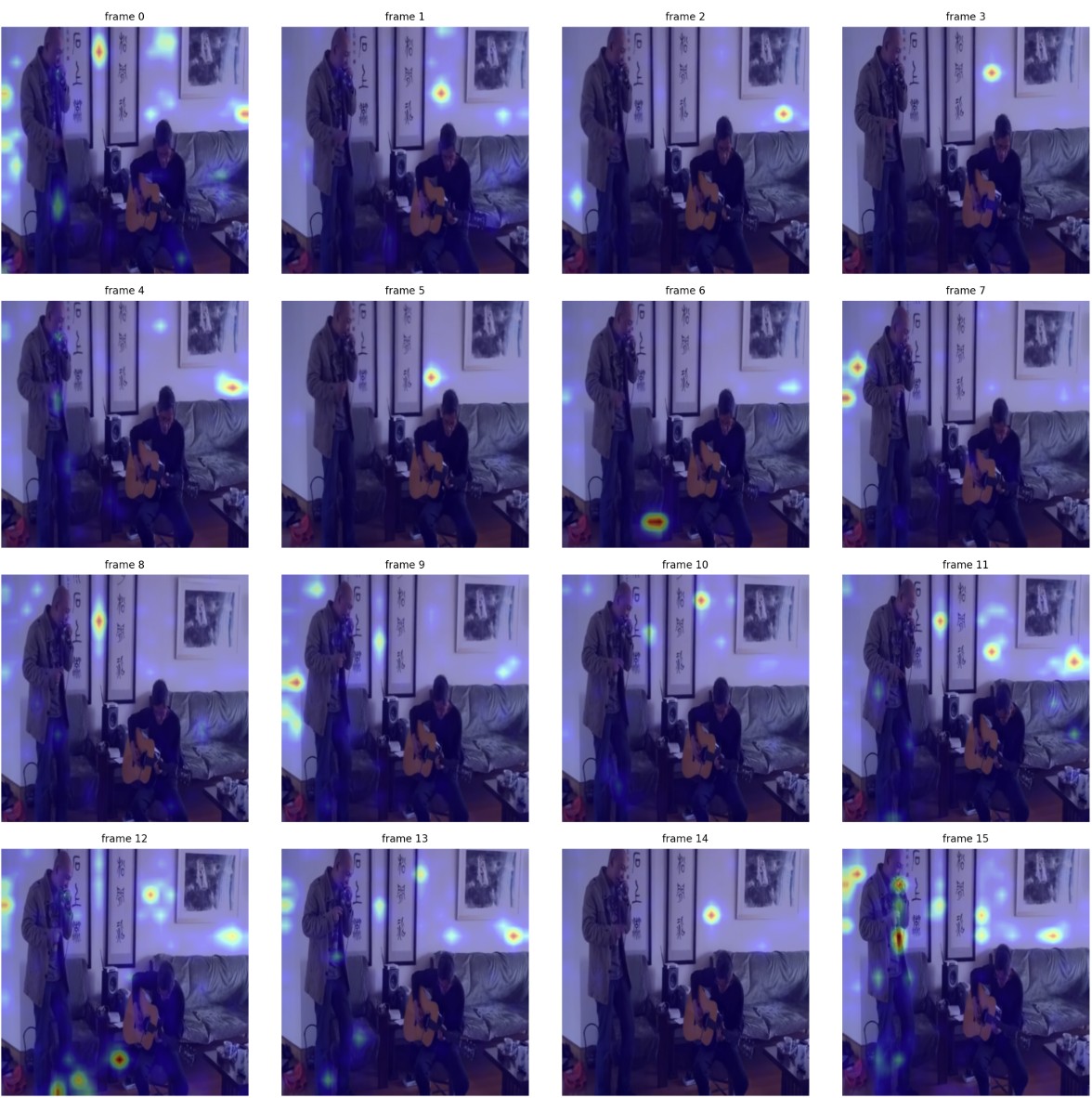}
    \caption{Gradient norm of the input tokens}
  \end{subfigure}
  \hfill
  \begin{subfigure}[t]{0.23\linewidth}
    \centering
    \includegraphics[width=\linewidth]{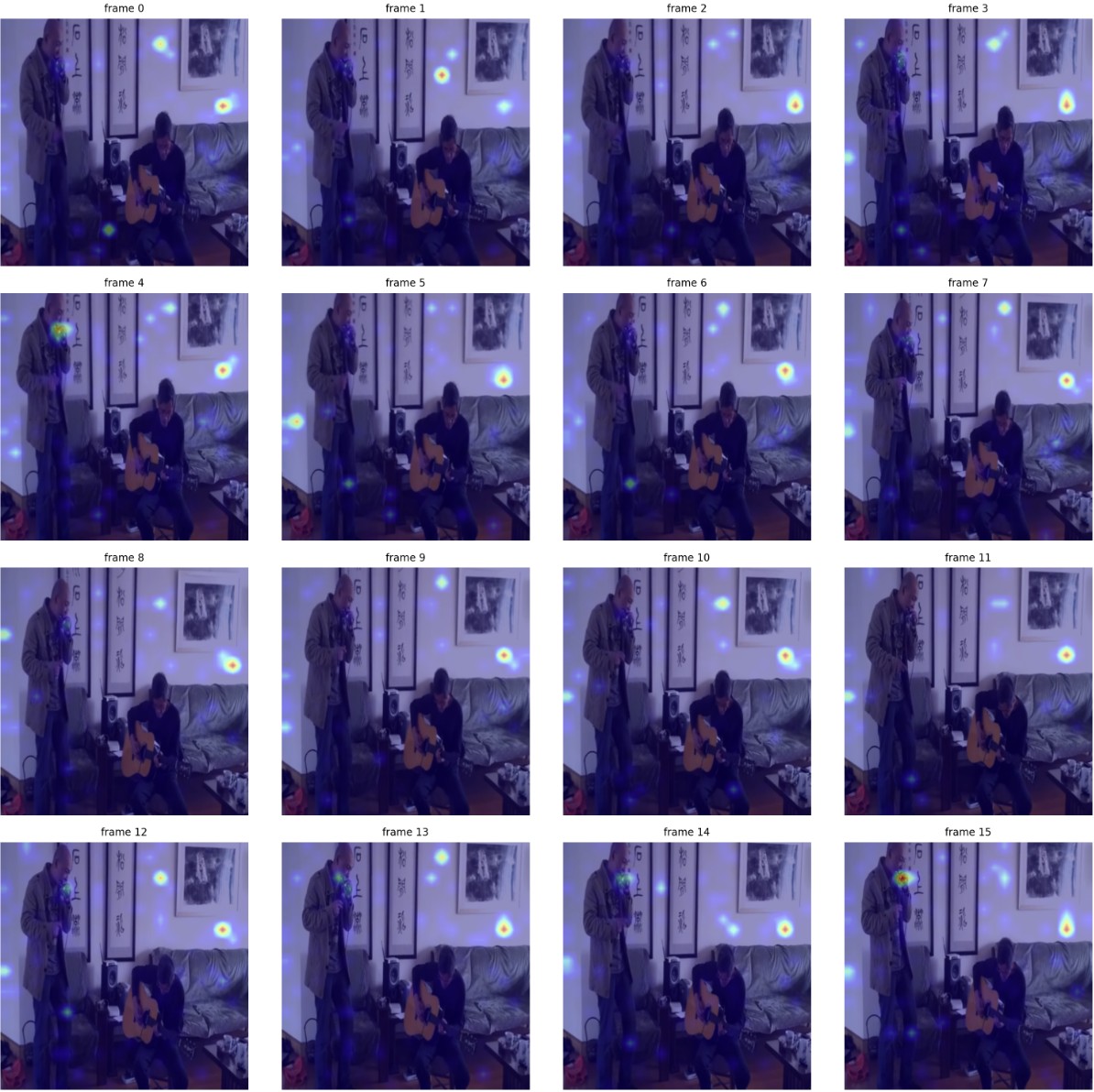}
    \caption{Gradient norm of the output tokens}
  \end{subfigure}
  \hfill
  \begin{subfigure}[t]{0.23\linewidth}
    \centering
    \includegraphics[width=\linewidth]{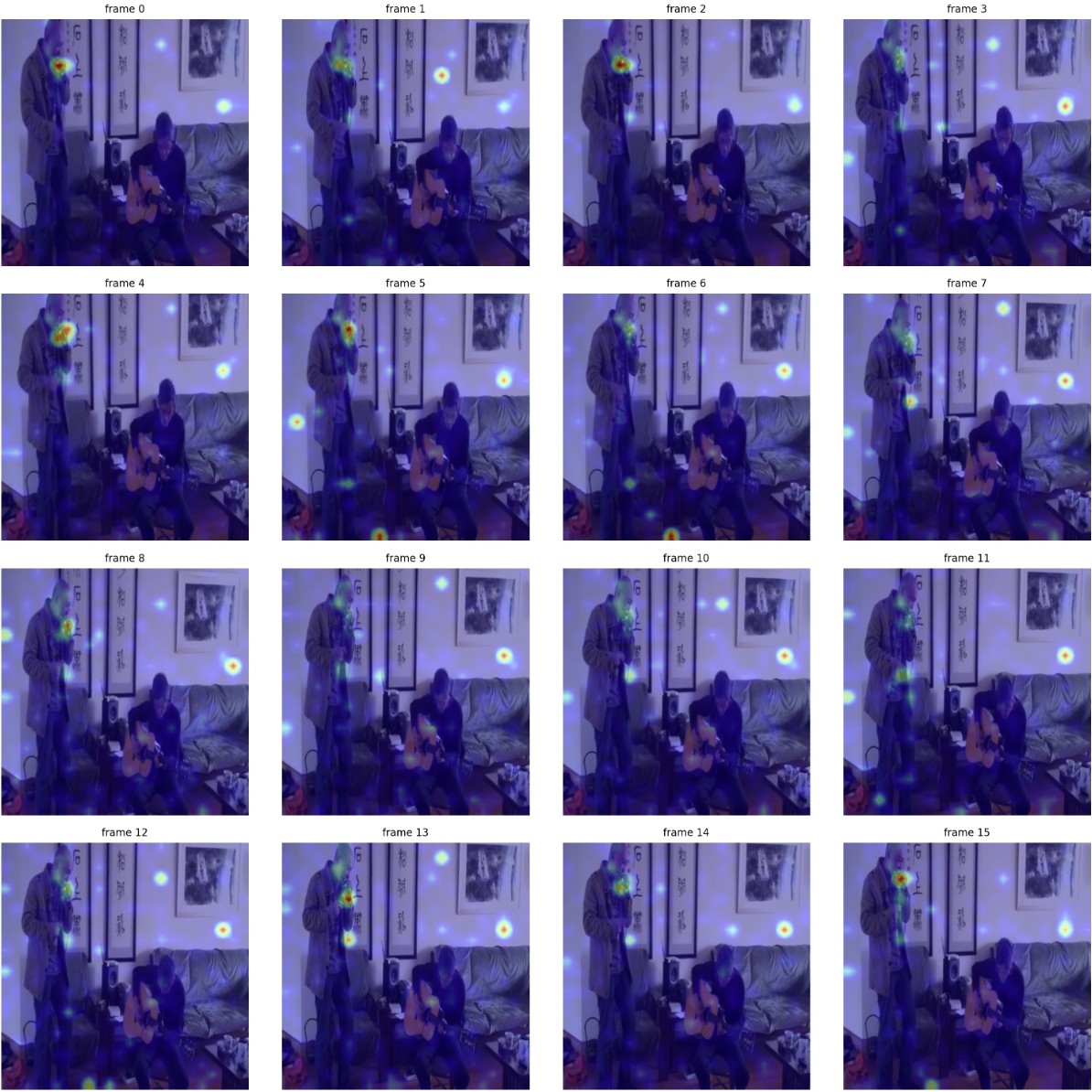}
    \caption{Attention maps of the original pe-av}
  \end{subfigure}
  \hfill
  \begin{subfigure}[t]{0.23\linewidth}
    \centering
    \includegraphics[width=\linewidth]{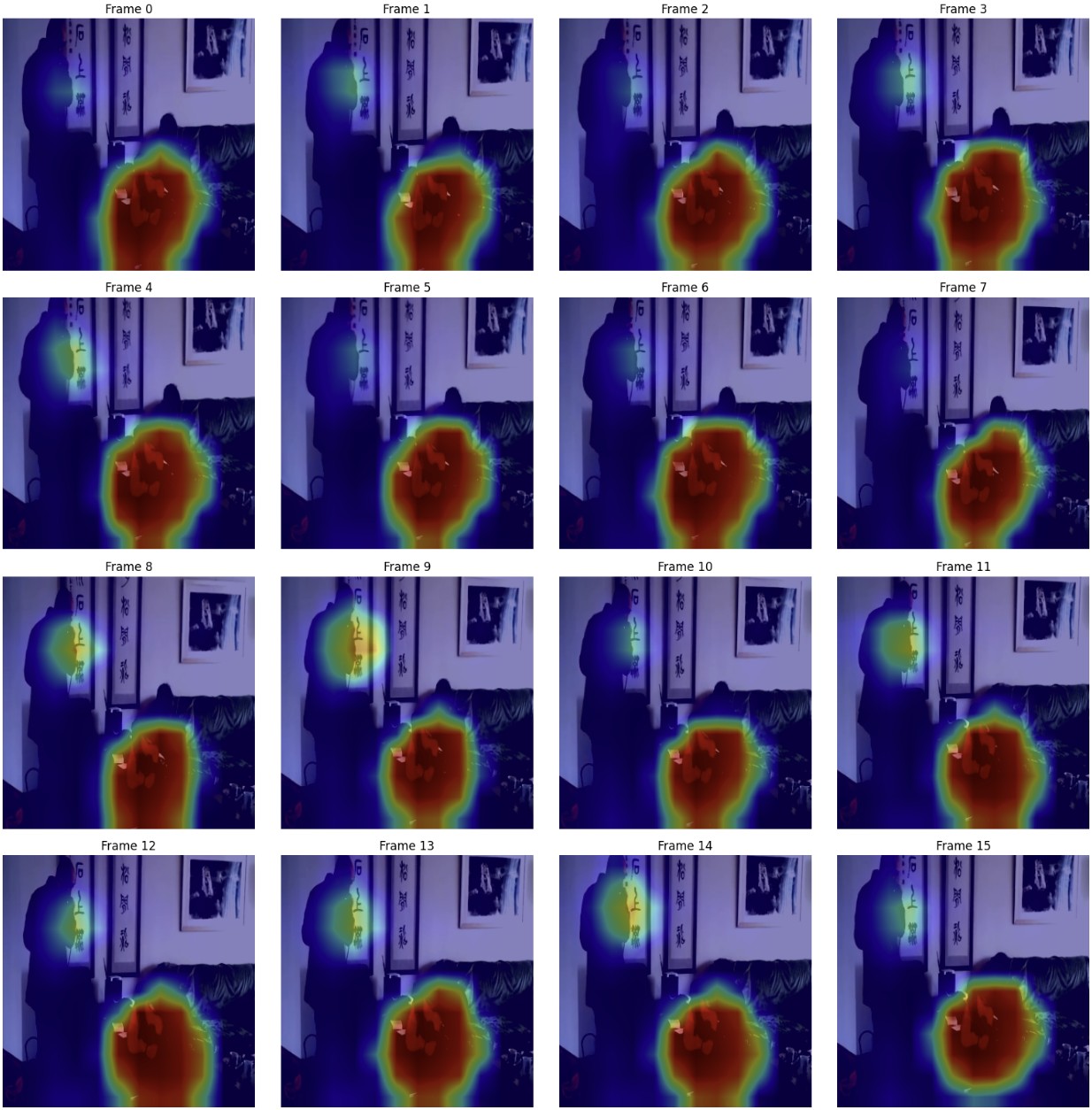}
    \caption{Attention maps of our model}
  \end{subfigure}
  
  \caption{PE-AV does not encode audio-visual spatial correspondences. The norm of the gradient of the input and output tokens of the frame encoder with respect to the audio visual similarity does not exhibit interpretable patterns. The attention of the original attention pooling of PE does not either. However our model focuses exactly on the sounding regions of the image.}
  \label{fig:peav_grad}
\end{figure}

Figure~\ref{fig:peav_grad} helps clarify why localization cannot be directly recovered from the original PE-AV model by showing segmentation obtained for 16 frames of the video using different methods relying on PE-AV. The gradient norms of the input and output visual tokens (of the frame encoder) with respect to the audio--visual similarity are diffuse and do not isolate a clear sounding region. Likewise, the ``attention maps of the original PE-AV'' should not be interpreted as localization maps: they correspond to the attention paid by the [CLS] token during the final attention-pooling step used to aggregate frame information before feeding it to the video encoder. Since this pooling is optimized for global retrieval rather than spatial grounding, its attention remains broad and mostly related to outlier tokens that encode most of the global information. In contrast, our AiSP maps are explicitly produced by using the audio token to query dense visual tokens, which makes them much more spatially specific.

\section{Additional qualitative study}
\begin{figure}[htbp]
\centering
\newcommand{\sqimg}[1]{%
\includegraphics[width=0.2\textwidth,height=0.2\textwidth,keepaspectratio=false]{#1}
}

\setlength{\tabcolsep}{2pt} 
\begin{tabular}{cccccc}
\textbf{Input} & \textbf{LAIP} & \textbf{Input}  & \textbf{LAIP} \\

\sqimg{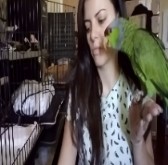} &
\sqimg{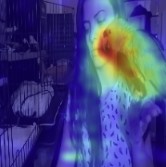} &
\sqimg{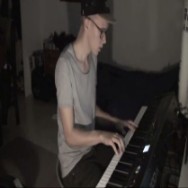} &
\sqimg{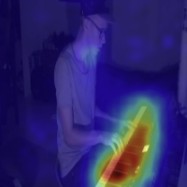} \\

\sqimg{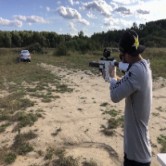} &
\sqimg{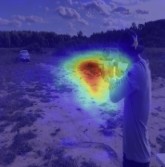} &
\sqimg{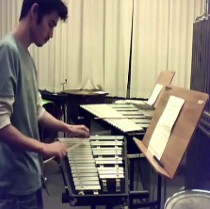} &
\sqimg{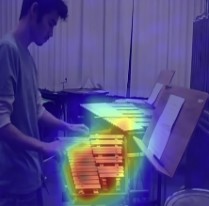} \\
\end{tabular}

\caption{Qualitative samples of LAIP on the MS3 (first row) an S4 (second raw) dataset.}
\label{fig:addQuali}
\end{figure}

Figure~\ref{fig:addQuali} shows some qualiative segmentation of LAIP on S4 and MS3 dataset. Our method localize accurately the sound sources, without outliers, even when there are multiple possible sound sources in the frame.

\begin{figure}[htbp]
\centering

\newcommand{\sqimg}[1]{%
\includegraphics[width=0.2\textwidth,height=0.2\textwidth,keepaspectratio=false]{#1}
}

\setlength{\tabcolsep}{2pt} 
\begin{tabular}{cccccc}
\textbf{Input} & \textbf{LAIP} & \textbf{Input}  & \textbf{LAIP} \\

\sqimg{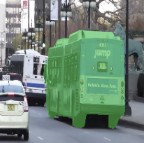} &
\sqimg{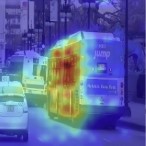} &
\sqimg{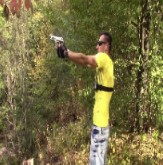} &
\sqimg{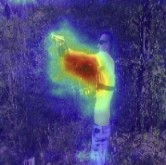} \\
\end{tabular}

\caption{Failure cases of LAIP.}
\label{fig:fail}

\end{figure}

Figure~\ref{fig:fail} shows a few failure cases of LAIP. When the source is very small or very big, the attention sometimes fails to segment it correctly.

\newpage

\end{document}